%% file: main.tex
\documentclass[sigconf]{acmart}

\AtBeginDocument{%
  }

\usepackage{xcolor}
\usepackage[skip=0.3\baselineskip]{caption}
\usepackage{fontawesome}
\usepackage{enumitem}
\usepackage{listings}
\usepackage{soul}
\usepackage{xspace}
\usepackage{multirow}
\usepackage{makecell}
\usepackage{tabularx}
\usepackage{array}
\usepackage{tikz} 
\usepackage{comment}

\newcommand{\system}{VideOrigami}
\newcommand{\pt}[1]{\textcolor[HTML]{3F91A0}{\textbf{\textit{\{#1\}}}}}
\newcommand{\nul}{\ensuremath{null}}

\copyrightyear{2025}
\acmYear{2025}
\setcopyright{cc}
\setcctype{by-nc}
\acmConference[CHI '25]{CHI Conference on Human Factors in Computing Systems}{April 26-May 1, 2025}{Yokohama, Japan}
\acmBooktitle{CHI Conference on Human Factors in Computing Systems (CHI '25), April 26-May 1, 2025, Yokohama, Japan}\acmDOI{10.1145/3706598.3713401}
\acmISBN{979-8-4007-1394-1/25/04}

\begin{document}

\title[Compositional Structures as Substrates for Human-AI Co-creation Environment]{Compositional Structures as Substrates for Human-AI Co-creation Environment: A Design Approach and A Case Study}

\author{Yining Cao}
\email{rimacyn@ucsd.edu}
\orcid{0000-0002-3962-2830}
\affiliation{%
  \institution{University of California San Diego}
  \city{La Jolla}
  \state{CA}
  \country{USA}
}

\author{Yiyi Huang}
\email{yih045@ucsd.edu}
\orcid{0009-0001-8941-8061}
\affiliation{%
  \institution{University of California San Diego}
  \city{La Jolla}
  \state{CA}
  \country{USA}
}

\author{Anh Truong}
\email{truong@adobe.com}
\orcid{0009-0005-5409-7287}
\affiliation{%
 \institution{Adobe Research}
  \city{New York}
  \state{NY}
  \country{USA}
}

\author{Hijung Valentina Shin}
\email{vshin@adobe.com}
\orcid{0000-0001-8798-4580}
\affiliation{%
  \institution{Adobe Research}
  \city{Cambridge}
  \state{MA}
  \country{USA}
}

\author{Haijun Xia}
\email{haijunxia@ucsd.edu}
\orcid{0000-0002-9425-0881}
\affiliation{%
  \institution{University of California San Diego}
  \city{La Jolla}
  \state{CA}
  \country{USA}
}

\renewcommand{\shortauthors}{Cao et al.}

\input{tex/0_abstract}



\begin{teaserfigure}
  \centering
   \includegraphics[width=1\textwidth]{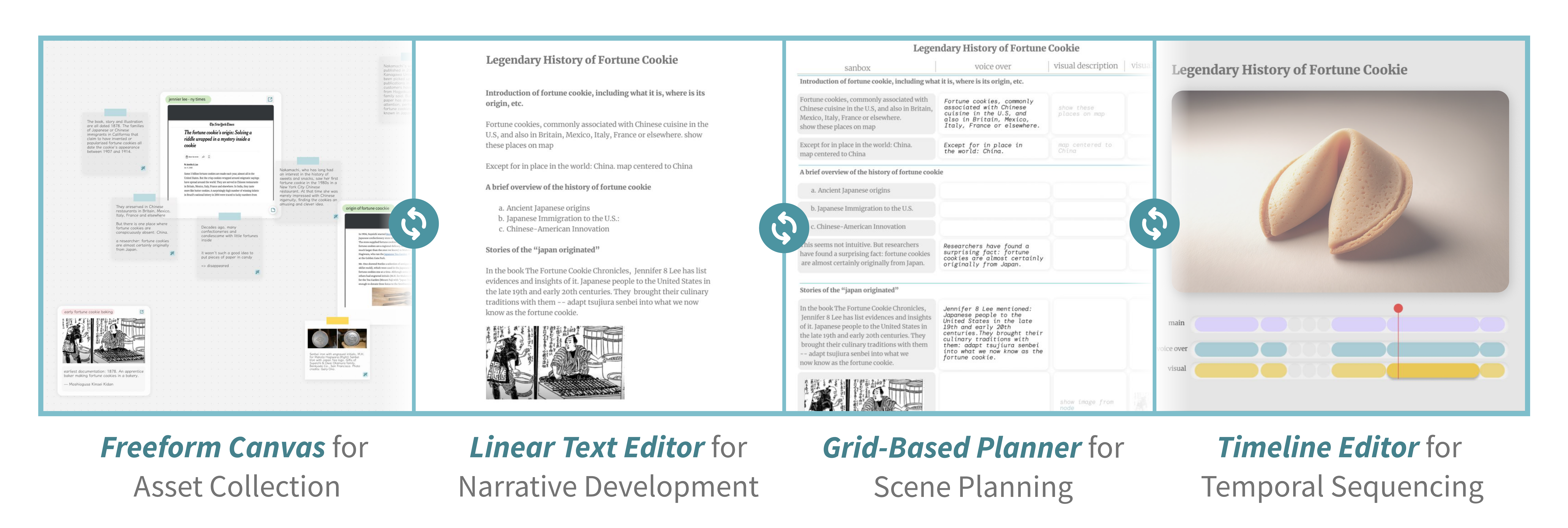}
 \caption{The user interface of \system{}, a human-AI video co-creation environment, developed using the proposed approach of combining compositional structures and AI. The compositional structures facilitate inspection and control of AI generation, and AI facilitates information transformation and synchronization within and across the structures.}
  \label{fig:teaser}
  \Description[]{ A teaser figure showing 4 images. The first image, captioned ``Asset Collection``, includes multiple PDFs and image previews on a canvas. One PDF preview is surrounded by multiple sticky notes. The second image, captioned ``Narrative Development``, includes formatted text and one image written into a linear structure. The third image, captioned ``Scene Planning``, includes a table with several columns (named sandbox, voice over, and visual description). The fourth image, captioned ``Video Editing``, includes a video editor with a video preview and timeline tracks. }
\end{teaserfigure}

\maketitle
\input{tex/01_intro}
\input{tex/02_relatedwork}
\input{tex/03_design}
\input{tex/04_formative}
\input{tex/05_ux}
\input{tex/06_evaluation}
\input{tex/07_discussion}
\input{tex/08_conclusion}

\bibliographystyle{ACM-Reference-Format}
\input{main.bbl}

\appendix
\input{tex/09_appendix}

\end{document}

%% file: tex/0_abstract.tex
\begin{abstract}
It has been increasingly recognized that effective human-AI co-creation requires more than prompts and results, but an environment with empowering structures that facilitate exploration, planning, iteration, as well as control and inspection of AI generation. Yet, a concrete design approach to such an environment has not been established. 
Our literature analysis highlights that compositional structures—which organize and visualize individual elements into meaningful wholes—are highly effective in granting creators control over the essential aspects of their content. However, efficiently aggregating and connecting these structures to support the full creation process remains challenging. We, therefore, propose a design approach of leveraging compositional structures as the substrates and infusing AI within and across these structures to enable a controlled and fluid creation process. We evaluate this approach through a case study of developing a video co-creation environment using this approach. User evaluation shows that such an environment allowed users to stay oriented in their creation activity, remain aware and in control of AI’s generation, and enable flexible human-AI collaborative workflows. 
\end{abstract}

%% file: tex/01_intro.tex
\section{Introduction}
\label{sec:intro}

Content creation is inherently iterative, involving exploration, planning, and refining. While advanced AI models are capable of generating high-quality text, images, and video clips from prompts \cite{GPT4TechReport, bommasani2021foundationmodels},  the HCI community has argued that relying on the prompt-generation paradigm alone is inefficient due to the lack of controllability and interpretability desired in the creative processes \cite{controlnet, whyjonnycantprompt}. Prior works have explored leveraging various external structures to augment human-AI collaboration: such as leveraging a chain structure to break down a complex task into smaller steps \cite{wu2022aichain}, employing content-specific structures to control AI generation (e.g., narrative structure in writing \cite{kim2023metaphorian, zhang2023visar}), and organizing generated text using diagrams and hierarchical structures for comprehension \cite{graphologue2023, sensecape2023}.

These prior works collectively suggest that effective human-AI co-creation of complex content (e.g., scientific writing, music, narrative video) goes beyond simple prompt-generation cycles, requiring an environment with empowering structures to ground AI generation,  facilitate human ideation, and support design iteration \cite{norman2014things, schon2017reflective, wu2022aichain, suh2024luminate}. 
Yet, there has not been an explicitly formulated approach to guide the development of such environments. Specifically, we lack systematic guidance on (1) what is the design process to follow, (2) what are the essential structures to consider, (3) how should AI be integrated with these structures, and (4) what are the benefits and challenges of such an environment. This work attempts to answer these questions. 

Toward this goal, we surveyed prior research that investigated challenges and developed systems to support creative activities in a variety of domains, including writing, multimedia posts, podcasts, music, and video production. From this analysis, we identified a common approach in designing interfaces for supporting various forms of content creation: the use of \textit{compositional structures}. 
We refer to compositional structures as structures that visualize and organize \textit{individual components} of content into a cohesive and meaningful whole based on specific \textit{content aspects}. 

Our analysis revealed that compositional structures address four key content aspects: spatial (e.g., layout in graphical design), temporal (e.g., pacing in videos), narrative (e.g., storytelling coherence), and congruent (e.g., integration of multimodal elements such as text, visuals, and audio).
For example, a narrative graph represents storylines as nodes and edges, enabling creators to inspect the flow of narrative points and experiment with alternatives; a multi-track timeline organizes individual video and audio clips along a temporal axis, supporting creators in sequencing, aligning, and adjusting the pacing of the clips. These structures not only assist in organizing and editing content by defining individual components and their organizational rules, but also provide functional affordances that guide creators through complex workflows, enabling efficient inspection, iteration, and refinement.
While these structures can be employed individually, complex creative processes often require multiple structures to interoperate. For example, in narrative video creation, a timeline may synchronize with a storyboard to ensure the alignment between visual sequences and story progression.

Informed by the literature analysis, we propose a design approach for developing human-AI co-creation environments, which consists of four steps
(1) identifying relevant compositional structures and their desired interconnections, (2) designing individual structures tailored to content aspects and workflow requirements, (3) aggregating these structures into a unified environment, and (4) infusing AI to support content creation and synchronization. With this approach, we aim to provide actionable guidance for building environments that balance human agency with AI augmentation, enabling effective human-AI co-creation.

An ideal evaluation of a design approach is to test it across multiple domains. This is challenging in terms of scope, as building an environment to support a single domain's extended workflow demands substantial design and development effort. Therefore, we opted for a case study in video creation. This domain encompasses the four content aspects identified in our literature analysis---spatial, temporal, narrative, and congruent---and thus warrants reasonable generalizability. We conducted a formative study to identify common compositional structures in video creation workflows as well as practices and challenges associated with these structures. We then developed a human-AI video co-creation environment, VideOrigami, by infusing these compositional structures with AI. We evaluated this co-creation environment by conducting a user evaluation with ten video creators. This study enabled us to investigate the benefits and challenges of such an environment and discover new creation patterns resulting from the reduced cost of aggregating compositional structures and integrating AI.   

Together, this work makes the following contributions.

\vspace{-5pt}
\begin{itemize}[leftmargin=*]
    \item \textbf{A literature analysis} of prior work across multiple creative domains, identifying compositional structures as a foundational design element for human-AI co-creation environments and summarizing the design practices, challenges and opportunities of leveraging compositional structures. 
    \item \textbf{A design approach} that proposes using the compositional structures as substrate for human-AI co-creation environment to ground the generation process; and infusing AI within and across these structures to enable flexible creation workflows.
    \item \textbf{A case study} where we developed a human-AI video co-creation environment, demonstrating the feasibility of instantiating the proposed approach.
    \item \textbf{A user evaluation} of the developed environment, validating the effectiveness of the approach and uncovering new patterns of human-AI collaboration enabled by the integration of compositional structures and AI.
\end{itemize}

%% file: tex/02_relatedwork.tex
\section{Related Work}
\label{sec:related}

Our research proposes a human-AI collaboration paradigm by grounding AI automation with compositional structures. We herein review prior work on human-AI collaboration and the role of compositional structures in scaffolding content creation. Additionally, we examine relevant literature on video creation.

\subsection{Supporting Human-AI Collaboration}

Recent advancements in generative AI are shifting content creation from relying on low-level manual editing to guiding AI with high-level instructions and goals~\cite{videoworldsimulators2024, team2023gemini, achiam2023gpt}. This is a significant step towards the human-AI symbiotic collaboration that Licklider envisioned~\cite{licklider1960man}. To support effective human-AI collaboration, recent work has made progress on various fronts, including design guidelines~\cite{mixedInitiativeUI, guidelinesHAI}, analytical frameworks~\cite{aialignment, gulfofenvisioning}, and interaction techniques~\cite{wei2022chain, graphologue2023}. 
For example, Amershi et al. proposed guidelines for designing human-AI interaction, which describe the desired high-level 
qualities of human-AI interfaces, such as ``show contextually relevant information'' and ``learn from user behavior''~\cite{guidelinesHAI}. Terry et al. and Subramonyam et al. proposed extending Norman’s Gulfs of Execution and Evaluation with Process Gulf~\cite{aialignment} and Envisioning Gulf ~\cite{gulfofenvisioning}, respectively, to better model human-AI interaction. 

Recognizing the lack of control and interpretability of the prompt-generation paradigm, prior work explored incorporating various structures into human-AI interaction, such as breaking down complex tasks into granular steps~\cite{wu2022aichain}, visualizing the generation space using key content dimensions~\cite{brade2023promptify, suh2024luminate}, adding additional visual structures (e.g, human poses) to control image generation~\cite{zhang2023adding}, using narrative structure~\cite{zhang2023visar}, node-link structures (e.g.,diagrams~\cite{graphologue2023}, and hierarchical spatial structures~\cite{sensecape2023}) to organize generated text. 

We extend the prior work on employing structures to improve human-AI interaction. Beyond utilizing individual structures for specific tasks, we explore how to develop co-creation environments enriched with these structures~\cite{norman2014things, schon2017reflective, kirsh2010thinking}. 
Buschek’s work on AI writing tools offers a related perspective, identifying an interface design pattern of collaging fragmented views to create the interfaces of writing systems~\cite{CollageNewWriting}. Our approach also draws on the concept of information environments built from ``information substrates" ~\cite{unifiedprinciple, klokmose2015webstrates, crosstalk, informationspaceFox}.
We propose employing compositional structures as a type of substrates, and infusing AI within and across them to create the human-AI co-creation environments.


\subsection{Effectiveness of Compositional Structures}

Compositional structures that describe the form, arrangement, and relationships of components have been found to be effective in facilitating the creation, consumption, evaluation, and iteration of information content~\cite{tableofcontentforvideo, reverseoutline, explodedview, cao2023dataparticles}.

Visualizing compositional structures can help users develop an overall understanding of the content and enable efficient consumption. For example, exploded-view drawings are effective in communicating the composition of various parts to inform the assembly, disassembly, and repair of mechanical components~\cite{explodedview}. The table of contents in books enables readers to quickly review the structure of a book and facilitate navigation. Similar structures have been adopted to assist in the consumption of videos by segmenting long-form videos into smaller and skimmable chunks based on narrative structures, such as SceneSkim~\cite{SceneSkim} and Video digests~\cite{videodigest}. 

During content creation, compositional structures enable creators to inspect and define the structures of the content and support structural revision~\cite{flower1981cognitive, kim2023metaphorian}. For example, exploring the composition of ideas during pre-writing was found effective in improving writing quality ~\cite{flower1981cognitive}, and reverse outlining can effectively aid in the structural revision of the documents~\cite{reverseoutline}. For music, compositional structures have been devised to facilitate the composition of chords~\cite{PaperTonnetz} and the entire music score~\cite{paperSubstrateForMusic}.  
For filmmaking, the three- or five-act narrative structures are established practices to create compelling narratives~\cite{field2005screenplay}. Inquiry-based structure is commonly used in science communication videos to maintain viewer engagement~\cite{billionsView}. Screenplay, storyboards, and audio-video scripts are commonly used to plan and develop films and videos~\cite{field2005screenplay, videoMosaic}. 

Given the effectiveness of compositional structures, recent work has sought to leverage them to assist human-AI co-creation. For example, research explored leveraging compositional structures to help writers plan, organize, and revise their writing with AI, such as using AI to generate outlines~\cite{ContinuousTextSummaries} or to generate passages based on argumentative structures~\cite{zhang2023visar}. Metaphorian supports creating extended metaphors by allowing users to define the structures of concepts they want to explain and leverage AI to generate sets of concepts exhibiting congruent structures~\cite{kim2023metaphorian}. These works, however, typically focus on utilizing a single compositional structure. In contrast, most content creation involves extended workflows that interleave many compositional structures in a highly dynamic and contingent manner~\cite{nardi1995studying, suchman1987plans}. Therefore, we explore how to develop human-AI co-creation environments with multiple compositional structures that support an entire creation workflow.

\subsection{Supporting Video Creation}

Video is a highly versatile medium that can integrate various forms of content, such as images, animations, text, infographics, sketches, sounds, and recordings. Composing these numerous heterogeneous materials of different modalities and formats into a coherent audio-visual piece is a highly challenging and tedious task, and therefore, received significant attention from HCI.  

Researchers and practitioners have proposed principles and practices regarding video composition, and developed many authoring support systems based on them. For example, the Congruence Principle states that the content and format of the visual content should be congruent to those indicated in the narrative~\cite{tversky2002animation}. Leveraging the desired congruence between the visual content and the underlying narrative, research has explored offloading the tedious, frame-level interactions of clips such as visual search~\cite{xia2020crosscast, huber2019b}, cut placement~\cite{truong2016quickcut, placingcuts}, and clip sequencing~\cite{wang2024lave, wang2019write, underscore, contentbased} with the synchronized manipulation of the corresponding scripts ~\cite{placingcuts, crosspower}. For example, Quickcut enables the automatic assembly of shots into a whole video by temporally aligning the shots with the video script~\cite{truong2016quickcut}. Crosspower leverages the semantic structures in the scripts to facilitate the spatial composition of visual materials in the scene~\cite{crosspower}.
Research has also explored leveraging the compositional structures of existing content in other media to generate videos. For example, end-to-end systems have been developed to create videos by transforming the composition of static content (e.g., documents, webpages), such as generating TikTok videos from news articles~\cite{wang2023reelframer} and marketing videos from websites~\cite{chi2020automatic}. 

We developed a human-AI video co-creation environment following our proposed design approach. As we will demonstrate, the interconnected and intelligent compositional structures serve as a versatile interface foundation that can coherently support many of the existing techniques but also afford new ones in the context of human-AI co-creation. Additionally, our user evaluations provide new insights into the challenges and opportunities in human-AI co-creation. The notorious complexity of video content and its workflow make video creation an ideal domain for validating the design approach, warranting considerable generalizability of the proposed approach and the resulting findings.

%% file: tex/03_design.tex
\section{Compositional Structures as Substrates: A Design Approach}

We grounded the development of the design approach in the existing literature dedicated to designing interactive systems for content-creation activities. Specifically, we sought to understand: \textit{how existing systems design compositional structures to support the development of the various aspects of the content.}
We analyzed work targeting diverse domains, including writing, music, podcasts, interactive media, and video. These domains allowed us to comprehensively cover textual, visual, audio, and interactive content.
For each domain, we selected survey,  study, and system articles as seed articles and utilized a snowball method to collect relevant literature.

For each article, we annotated the compositional structures that were studied, along with the content aspects they aim to support and their functionality. We extracted excerpts from the papers regarding the design decisions and challenges associated with the structures. We stopped the snowball process when no new compositional structures were found. 
In cases where multiple articles addressed the same compositional structure, we prioritized those that were widely cited. Ultimately, we identified 55 papers, and more details of these papers are included in the Appendix~\ref{sec:apx-lit}.

\subsection{Definition of Compositional Structures}
\label{subsec:def}
Our literature analysis revealed that interfaces for supporting content creation are typically designed with one or more structures. These structures define the individual components and organizational rules for assembling the components based on specific aspects of the content, offering corresponding functional affordances that facilitate the composition process -- we refer to these structures as \textbf{compositional structures}.  By analogy with biological substrates—surfaces on which living organisms grow—we conceptualize compositional structures as substrates on which informational content `grows'.

We herein summarize the \textbf{utility of compositional structures} across key aspects of content creation, analyzing the design of their individual components, organizational rules, and functional affordances implemented in existing systems (Section~\ref{subsec:utility}). 
Among the reviewed papers, 47 out of 55 combined multiple compositional structures to support the creation process. By analyzing the \textbf{interleaving usage of different compositional structures}, we summarize current design practices for establishing synchronization of these structures and formalize the solutions for aggregating them into one workspace. Additionally, we highlight challenges identified in existing research and propose opportunities to leverage AI to support the iterative creation processes across multiple compositional structures (Section~\ref{subsec:interleave}).
This analysis provides insights into \textit{what} compositional structures should be integrated into co-creation environments and \textit{how} to infuse AI within and across structures to support creative workflows.

\begin{table*}[t]
\centering
\caption{Compositional Structures for Four Content Aspects with Example Systems from Literature Analysis}
    \includegraphics[width=\textwidth]{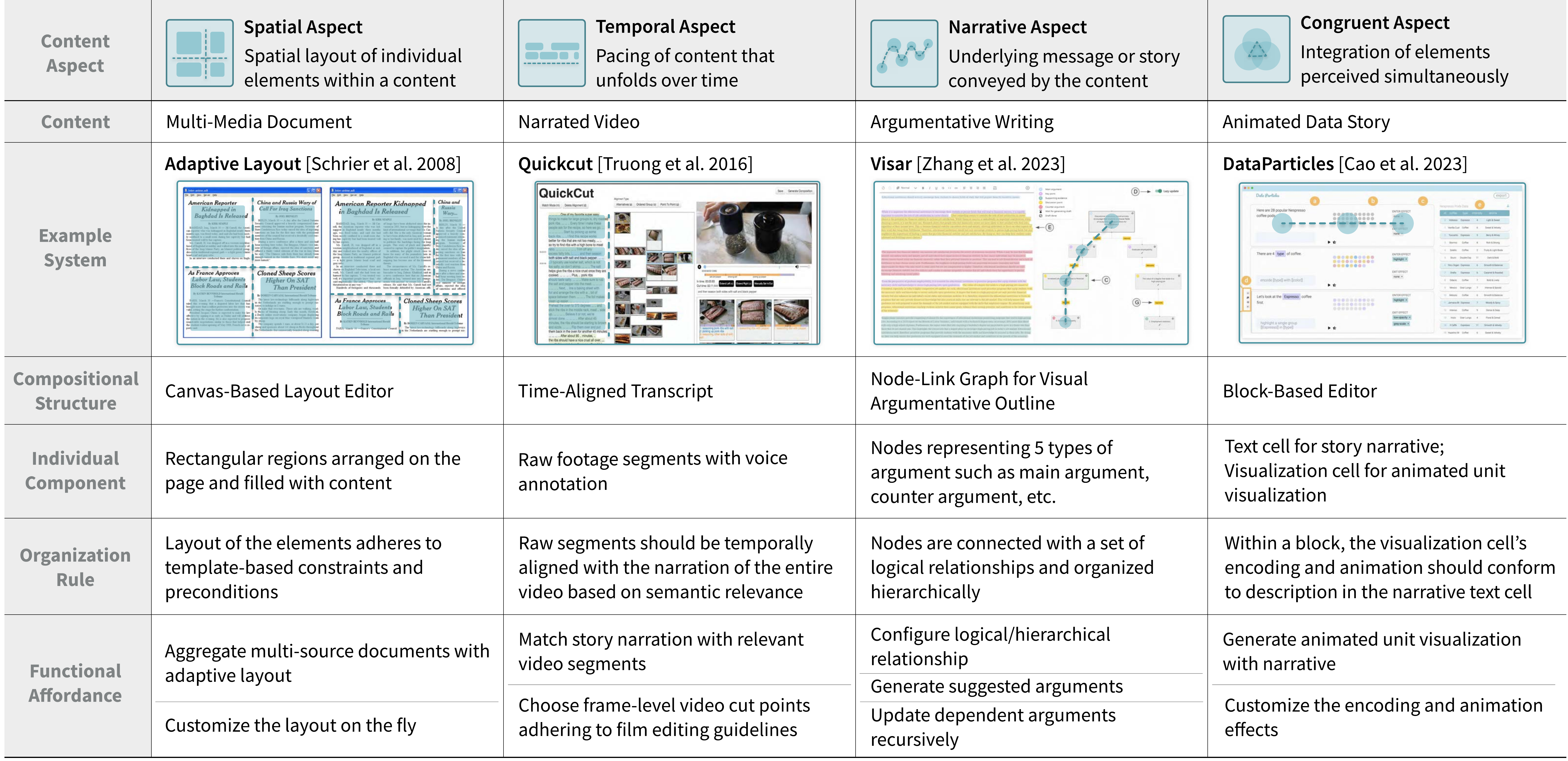}
    \label{fig:approach-comp1}
    \Description[]{A table describing compositional structures supporting each of the four content aspects: Spatial, Temporal, Narrative and Congruent. The table gives two exemplar structures for each content aspect. Each compositional structure is represented as a table row with columns detailing 1. a system in which it was leveraged and the domain of that system 2. the individual component composed by the structure, 3. organization rules that the structure enforces, and 4. the structure's functional affordances.}
\end{table*}

\subsection{Utility of Compositional Structures} 
\label{subsec:utility}
Each compositional structure assists in the creation of one or more aspects of the content. We categorized four key aspects: spatial, temporal, narrative, and congruent. 

\subsubsection{Spatial Aspect} 
This aspect primarily refers to the layout of individual components in the final content. It pertains to the organization of different elements in visual content for effective communication and appeal. Compositional structures supporting this aspect are often based on a \textit{free-form canvas} or \textit{grid system} with directly manipulable elements, allowing flexible adjustments of their position and size. The organizational rules in these structures aim to satisfy constraints and preferences specific to the creation context, such as design guidelines, stylistic choices, or screen sizes. The compositional structures reify these rules to enable easy reuse and adaptation by both creators and automated processes, including auto-suggested templates or computational layout adaptations.

\subsubsection{Temporal Aspect} 
This aspect addresses the pacing of content that unfolds over time. Interfaces designed to support this aspect typically feature a \textit{multi-track timeline} that allows creators to place elements along a time axis, define keyframes, and preview the progression. While pacing is critical for effective storytelling, most systems require the creators to manually achieve the ideal pacing, such as deciding the duration of shots and adding pauses to enhance narrative impact. Prior work has explored implementing organizational rules based on desired dialogue styles on pacing~\cite{computationalnotebookforcollaboration}. 
Other works explored supporting element alignment by time, such as synchronizing video segments with transcripts~\cite{truong2016quickcut, cao2024elastica, huber2019b}.
Such synchronization embedded in the timeline accelerates the creation process by alleviating the manual coordination efforts otherwise required to deal with different tracks in the timeline structure. 

\subsubsection{Narrative Aspect} 
This aspect refers to the underlying story conveyed by the content. Creators often need to experiment with different narratives to develop the most effective one. Compositional structures supporting this aspect vary in levels of abstractions: from conceptual elements like the luckiness of a character in storytelling~\cite{talebrush} to argumentative outlines in scientific writing~\cite{zhang2023visar}.
The organization rules and functional affordances of these structures depend on both the content types and the creative workflow. For example, \textit{graph structures} on a freeform canvas are commonly used for ideation, with association functions to generate or connect related concepts for brainstorming.
\textit{Hierarchical linear structures} are effective for planning and refinement, with summarization and expansion functions, such as summarizing paragraphs into a reverse outline or expanding simple headings into narrative points.           

\subsubsection{Congruent Aspect} 
This aspect addresses the integration of multiple individual elements perceived simultaneously during content consumption, such as notes in a musical chord~\cite{PaperTonnetz},
different modalities in a video~\cite{truong2016quickcut, videoMosaic}, and data visualizations and accompanying narratives in data storytelling~\cite{cao2023dataparticles}. To support this, compositional structures often associate these elements within a container to support reasoning: \textit{lines} connecting notes in a musical chord~\cite{PaperTonnetz}, \textit{storyboard cards} associating visuals with text~\cite{videoMosaic}, and \textit{blocks} combining visualizations and descriptions side-by-side~\cite{cao2023dataparticles}. These containers frequently incorporate auto-update functionalities to ensure synchronized updates among elements and may include functionalities for suggesting alternatives or auto-completing based on congruence rules. Furthermore, by bundling elements, creators can manipulate the container as a whole without compromising its internal congruence. Therefore, such structures often integrate congruence with other aspects, like narrative: for instance, a two-column structure, commonly used in video creation, expresses congruence within individual rows while supporting narrative progression through the linear arrangement of rows in a table format.

Notably, a single compositional structure can support multiple aspects simultaneously. For example, a transcript-based timeline supports both the temporal and narrative aspects as they have strong correspondence in certain video content. Similarly, the congruent aspect is often intertwined with others, as it inherently involves relationships between elements of different modalities. It is also important to acknowledge that while the spatial, temporal, narrative, and congruent aspects are prominent in the literature we have reviewed, they do not represent an exhaustive set of content aspects. For example, emotional resonance or interactivity may constitute additional dimensions that require further exploration.

\subsection{Interleaving Usage of Compositional Structures and Challenges}
\label{subsec:interleave}

Content creation is often multifaceted and highly iterative~\cite{challengeinmusicscorewriting, cao2023dataparticles}. These structures, however, are typically distributed across separate applications, resulting in fragmented workflows~\cite{Passage, cao2023dataparticles, crossdata}. 
From the papers that utilized multiple compositional structures, we summarized two primary needs:
\textit{inspecting different content aspects} and \textit{facilitating iterative transitions}. To address these needs and mitigate the challenges of fragmented workflows, a common design approach is to consolidate essential compositional structures into a single interface, and establish synchronization among them.
To inform our design approach, we summarize existing practices for integrating compositional structures and achieving synchronization across these structures.

As each compositional structure defines its individual components and their organization rules, the synchronization aims to establish connections between the components across structures while satisfying their respective rules. We summarized three aspects to consider for establishing synchronization and highlighted the design challenges within each aspect. 

\subsubsection{Defining Correspondences Between Individual Units Across Structures}
Establishing how individual components in one structure correspond to those in another is fundamental to achieving synchronization across structures. For example, a node in the narrative graph corresponds to a sentence in the text editor~\cite{zhang2023visar, talebrush}. Predefined correspondences help users understand the system’s logic and how modifications in one structure influence others. However, they can also limit flexibility. As the creative process evolves, these correspondences may become ambiguous, requiring reevaluation or reconfiguration, which can disrupt the creative flow.

\subsubsection{Determining Appropriate Synchronization Techniques}
We identified two common techniques for cross-structure synchronization: \textit{Synchronized Highlighting}, which highlights corresponding units across structures to aid navigation and reference, and \textit{Synchronized Editing}, where updates in one structure propagate to corresponding units in another. Systems with synchronized editing typically also include synchronized highlighting. The complexity of synchronized editing depends on how content is represented across structures. It may involve direct content transformation (e.g., transferring a phrase in an outline to a section heading), attribute-based transformation (e.g., a text snippet converted to a timeline duration), or advanced AI-driven generation (e.g., a paragraph converted to an outline point). While advanced transformations can streamline workflows, they often require significant review and adjustments. Simple synchronized highlighting may be preferable when creators prioritize manual control over automation.

\subsubsection{Configuring Updating Mechanism}
Two key factors influence the updating mechanism: (1) Directionality: updates can be \textit{bi-directional or one-directional}, where changes flow between structures in both directions or only from one to another; (2) Control: whether updates should be \textit{automated or user-controlled}, with automated updates ensuring consistency but potentially causing unintended overwrites, while user-controlled updates provide greater control but require additional actions that might be laborious. While most systems favor automated bi-directional updates for ease of use, this approach can pose challenges in preserving intentional modifications of content within one structure.

\subsection{Proposed Design Approach}
\label{subsec:designapproach}

Existing research demonstrates the utility of compositional structures and explores practices for constructing workspaces that leverage them. Insights from this literature inform key considerations for designing such structures and integrating AI functionalities within and across them. Based on these findings, we propose the following four-step design approach:

\begin{itemize}[leftmargin=2.5em]
\item [\textbf{ST1}]  \textbf{Identifying Compositional Structures and Their Desired Interconnections of a Creation Activity. }
The first step is to investigate the creation workflow for the targeted content. This includes identifying specific compositional structures used to address different content aspects, their roles at various workflow stages (e.g., ideation, editing, integration), and how these structures should interconnect to support transitions between them. 

 \item [\textbf{ST2}]  \textbf{Designing Individual Structures with Content Aspects and Workflow Requirements.}
For each identified structure, specify the individual components creators will manipulate and define the organizational rules (e.g., hierarchical relationships, temporal sequencing). The design of these structures should align with the intended creative outcomes and support efficient manipulation and arrangement within the structures.

\item [\textbf{ST3}] \textbf{Aggregating the Compositional Structures as the Foundation for the Co-creation Environment.}
The defined structures need to be integrated into a workspace with desired synchronization by defining corresponding units, synchronization techniques (e.g., synchronized highlighting or editing), and updating mechanisms (e.g., one or bi-directional).

\item [\textbf{ST4}] \textbf{Infusing AI Functionalities within and across Compositional Structures to Facilitate Content Creation and Synchronization.}
Based on the challenges within the creation workflow, automation needs to be infused within and across the structures. Within structures, AI should facilitate creating individual units adhering to their organizational rules; across structures, AI should maintain context awareness by managing references, coordinating interconnected content, and ensuring synchronization through appropriate updating mechanisms.
\end{itemize}

%% file: tex/04_formative.tex
\section{Understanding the Use of Compositional Structures in Video Creation}
\label{sec:formative}
We evaluate our design approach by executing the proposed steps. We begin with identifying the compositional structures and the desired interconnections. As indicated in the previous section, many prior works have been conducted to understand and alleviate the challenges in video editing \cite{chi2022synthesis, truong2016quickcut, xia2020crosscast}. However, there lacks work that specifically investigates compositional structures in the entire video creation workflows, especially the desired interconnections among them. Therefore, we conducted an interview study with video creators to holistically understand their creation processes.

\begin{figure*}[t]
\centering
\includegraphics[width=0.9\textwidth]{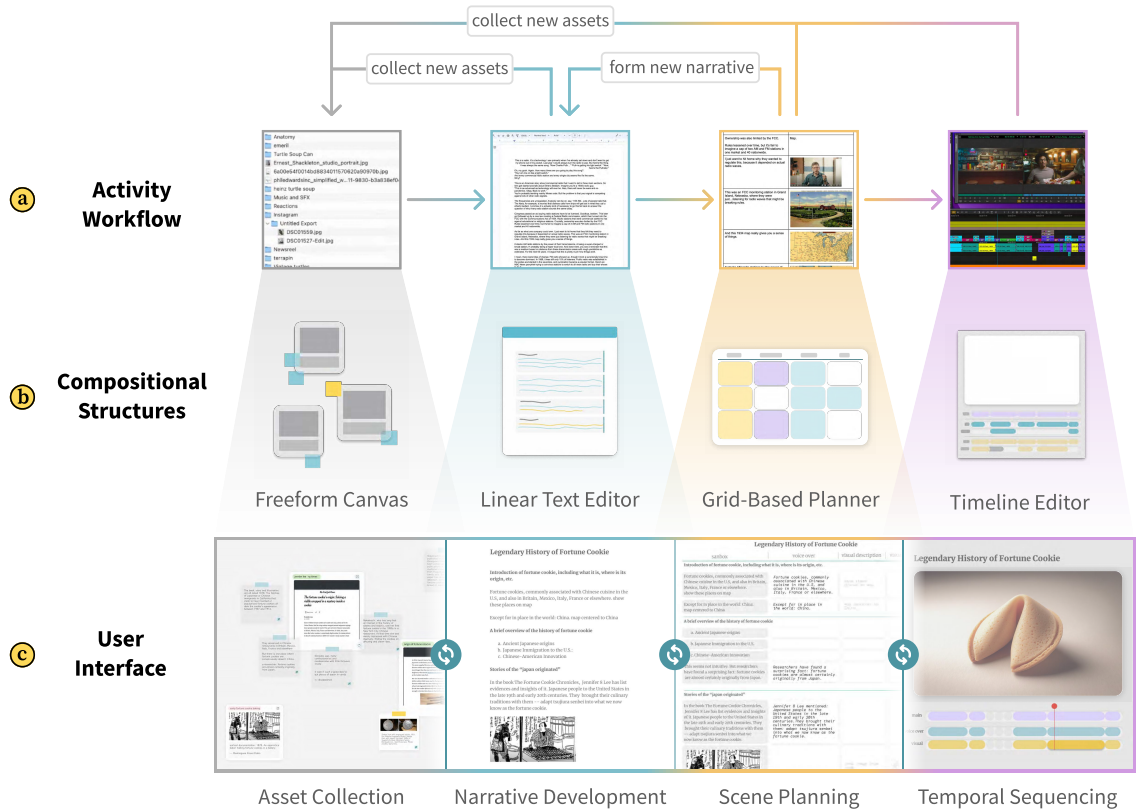}
\caption{Mappings between the compositional structures identified in the video creation workflow and the \system's user interface. (a) Workflow revolving around the compositional structures; (b) the underlying compositional structures; (c) four views in \system's user interface maps to a corresponding composition structure}
\label{fig:formativeandUI}
\Description[]{Three rows represent each mapping, as described in the main text. The first row labeled (a) shows a file directory, a word document editor, a table with one column of text and another column of images, and a traditional video editor. The second row (b) shows icons labeled `Canvas`, `Editor`, `Grid`, and `Timeline`. The third row (c) is the same as the teaser image. }
\end{figure*}


\subsection{Participants and Procedure}
\label{subsec:participants}
We interviewed five expert video creators, each with over five years of experience in video production and publishing. To ground the interviews in concrete examples, we asked the creators to share videos they had produced, along with any materials used for planning and prototyping. Participants were purposefully selected to ensure their videos covered diverse types of content. In total, the interviews covered 14 videos: 3 vlogs, 3 short films, 3 explainers, 3 video essays, and 2 animated films. The interviews began with questions regarding creators' professional experiences, followed by an in-depth discussion of the creation processes. Conducted via video calls, each interview lasted around 90 minutes.


\subsection{Compositional Structures in Existing Workflow}
\label{sec: formative-structures}
Despite the diverse video types, creators' workflows revolve around four key compositional structures shown in Fig.~\ref{fig:formativeandUI}b. We first review these structures and then dive into the challenges of working with these structures.  

\subsubsection{Ideation and Asset Organization with Freeform Canvas} 
\textit{``You could see there is no discipline here, because I am just throwing them all in.''} (E2). 
All creators use dedicated spaces to organize relevant assets, such as documents, videos, and images. 
As creators sift through the assets, they jot down associated notes, develop narrative points, and connect them to form the storylines.
Freeform canvases, such as Milanote (E1) and Miro board (E3, E5), were useful as they not only serve as the asset repository but also an ideation space where creators can arrange all materials to develop an understanding of the story. Creators frequently revisit this space to re-contextualize themselves with the materials.

\subsubsection{Narrative Development with Linear Text Editor}
\textit{``I just write out something almost off the top of my head with an idea.''} (E2)
Creators need a space to organize disconnected ideas into a cohesive storyline. At this stage, a linear structure can be helpful, as it ``actually helps me thinking.'' (E1)
This stage involves substantial iteration and engagement with content of mixed fidelity. 
Creators usually begin by putting down ideas and talking points that emerge during ideation and asset collection. The initial content could be ``written partially in full text, partially in outlining text.'' (E3) and as ``a combination of scripts and visuals''(E2, E3, E5).  Creators navigate through the content of mixed fidelity and modality and iteratively mold it toward a final storyline. 

\subsubsection{Scene Planning with Grid-Based Editor} 
\textit{``I need to put everything together, and see whether the visual goes well with the text, whether the temporal sequence makes sense.''}(E1)
Creators develop scene structures by examining the narrative and congruence aspects simultaneously: they experiment and specify the arrangement of materials to ensure visual and temporal coherence both within and across the scenes. A variety of grid-based structures are utilized, such as two-column scripts (E2, E5) and storyboards (E1, E4). For example, two-column scripts allow creators to arrange the narrative sequence in rows and organize the materials within a scene in columns (e.g., voiceover and desired visuals). 
This stage often features a ``random filling'' pattern, as creators may have incomplete ideas or uncertainty in certain parts of a scene, such as knowing the visual but not the voiceover, or vice versa. The grid structure also provides a clear overview of the creation progress---unfilled cells serve as visual indicators of incomplete elements (E2, E5).

\subsubsection{Spatial/Temporal Arrangement and Preview using Timeline-Based Editor} 
\textit{``You have been creating it as a creator, and now you are watching it as a viewer when putting them together.''} (E5) 
The timeline structure provides creators with fine-grained controls to refine the temporal sequence of the video, which can range from small pacing adjustments like trimming a clip to changing the clip sequences (E1, E2, E5). All timeline-based editors also provide controls for adjusting the placements of visuals for a selected time frame. At this stage, creators constantly assess the effectiveness of the storytelling by previewing the assembled parts. This iterative process of a perceptual reasoning is critical, as E3 explains: \textit{``Only when you see it do you realize it's not what you imagined.''}.


\subsection{Desired Interconnections Across Structures}
\label{sec:formal-interconnecton}
The current digital environment discourages iteration across structures, as the lack of synchronization between them often results in a high cost of context switching. This often incentivizes creators to confine their iterations to the single structure they are currently working in, rather than leveraging the one that would be most effective for the task.
For example, all creators we interviewed noted that once they transition to working on the timeline, they ``never go back''(E1), even when certain structural changes could benefit from using the narrative editor or scene planner (E2). Below, we summarize a set of cross-structure interconnections desired in the video creation workflows. 

\subsubsection{Collect and Refer to Materials Anytime} 
Creators often need to cross-reference assets while working on different structures at any workflow stage. E2 highlighted the frustration of repeatedly searching through scattered file folders to locate assets, which are often organized differently depending on tasks such as writing or timeline editing. They also mentioned that multiple passes are needed to ensure no assets are overlooked, consuming up to half a day. Systems should provide a centralized collection of assets, allowing creators to freely add assets within any structure and easily reference them when working on different structures.

\subsubsection{Develop Cohesive Narrative from Fragmented Notes} During asset collection, creators often generate fragmented ideas or notes that do not immediately fit into the narrative. At times, they may struggle to incorporate some interesting excerpts into the existing storyline (E3, E4). This suggests that systems should facilitate the quick integration of fragmented ideas into the narrative while ensuring narrative coherence.

\subsubsection{Provide a Warm Start for Developing the Scene Structure}
To develop the scene structure, creators often need to manually locate and transfer many materials into the appropriate categories (e.g., specific rows or columns), such as copy-pasting paragraphs from the narrative to different cells and inserting images and clips. This ``cold start'' process can be tedious and discourage them from using the scene structures (E1). Systems should automate the initial organization of content into scene structures, providing a ``warm start'' that reduces manual effort and streamlines the workflow.

\subsubsection{Enable Granular and Structural Adjustments for Temporal Sequencing} 
When adjusting pacing, creators often have diverse needs for corresponding edits: they require fine-grained controls, such as precise timing adjustments, which should be reflected across other structures to ensure these structures remain reusable for relevant tasks; they may also make broader structural changes to the sequence, which should be supported in alternative structures and automatically synchronized with the timeline. Systems should accommodate both granular and structural adjustments, ensuring synchronization across all relevant compositional structures.

%% file: tex/05_ux.tex
\section{\system{}: a Human-AI Video Co-Creation Environment} 
\label{sec:ux}
We develop \system{} by following the proposed design process, including surfacing the compositional structures in one unified space [ST2], 
defining their synchronization [ST3], and infusing AI within and across structures [ST4]. Section~\ref{sec:ux-within} explains the individual structures and AI-driven generation to support their completion; Section~\ref{sec:ux-across} covers synchronization across the structures and AI implementation; and Section~\ref{sec:ux-walkthrough} presents a scenario illustrating the creation of a video within the co-creation workspace ~\system{}.

\subsection{Compositional Structures and Within-Structure Generations}
\label{sec:ux-within}
We formally define the four compositional structures we identified: Freeform Canvas, Narrative Editor, Grid-based Scene Planner, and Timeline Editor (Fig. ~\ref{fig:formativeandUI}c)---along with their individual components, organization rules, and functional affordances. 

\begin{figure}[h]
\centering
\includegraphics[width=0.95\linewidth]{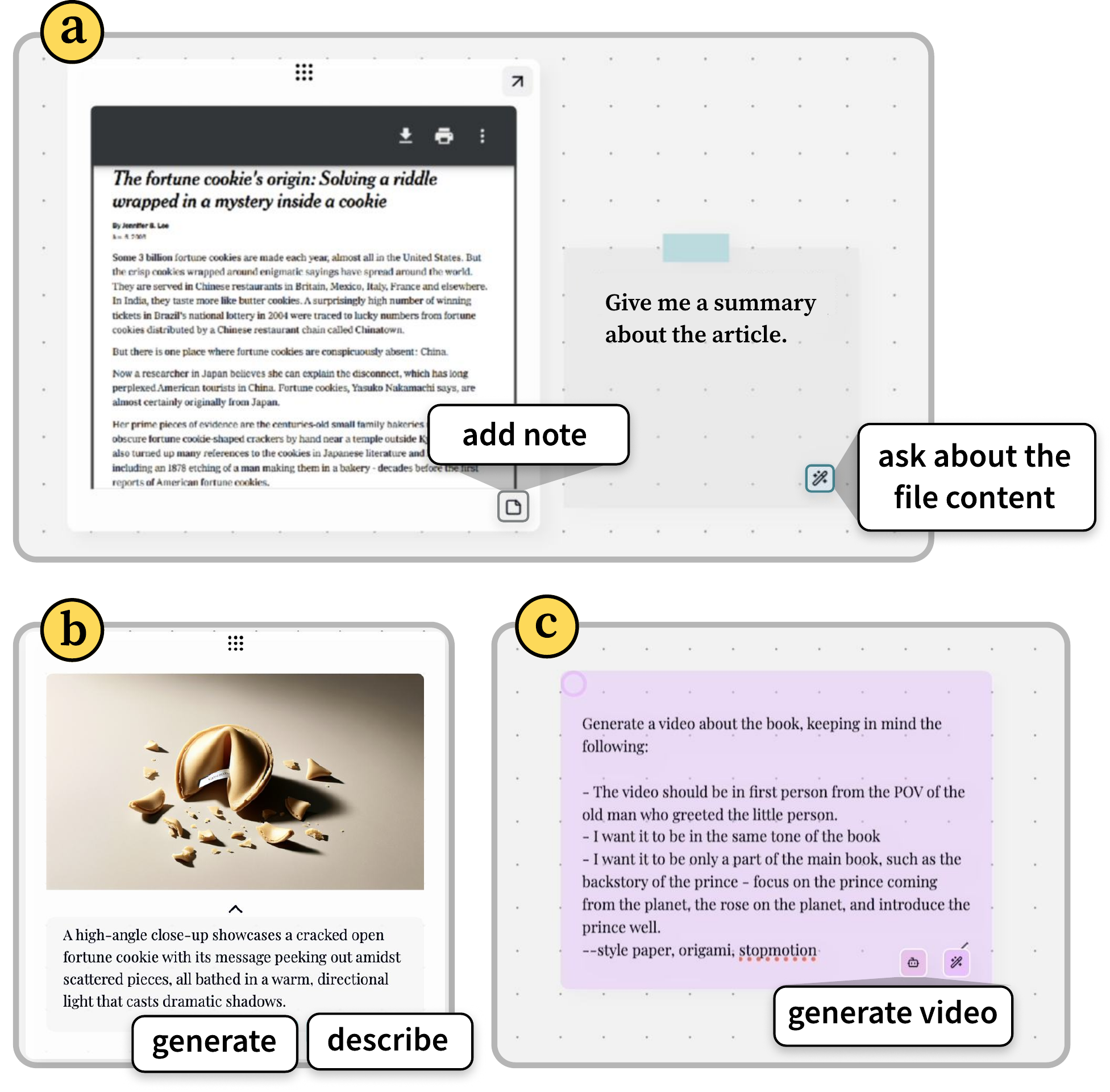}
\caption{Nodes supported in the freeform canvas structure.}
\label{fig:Canvas}
\Description[]{Three images are labeled (a), (b), (c). Image (a) shows a PDF preview in a square node with a button at the lower right that is highlighted as `add note`. To the right of the PDF preview is a sticky note with a button highlighted `extract content`. The second image (b) shows a square node with an image preview and description. Below are two buttons `regenerate` and `describe it`. In (c) shows a square node with text and two buttons, highlighted as `generate video` and `prompt to GPT`. }
\end{figure}

\textbf{The Freeform Canvas} leverages spatial organization to facilitate asset collection and exploration. 
\system{}'s implementation supports three types of nodes: \textit{Asset Nodes} (Fig.~\ref{fig:Canvas}a) for uploading media, importing web content, or generating visual content (Fig.~\ref{fig:Canvas}b). Each asset node has associated \textit{Note Nodes}, enabling creators to record information related to the assets when making sense of them. Users can manually edit notes or generate them using queries to extract specific content from the assets. The \textit{Prompt Nodes} (Fig. ~\ref{fig:Canvas}c) are used for generating certain parts of the video (as further elaborated in Section~\ref{sec:ux-walkthrough}).

\textbf{The Narrative Editor} that we utilize is a linear block-based text editor that consists of two block types: section blocks, which define high-level section headings to guide the overarching structure of the video, and paragraph blocks, which contain individual talking points within each section. Users can manually edit each block and use LLMs to generate section headings to outline the narrative or talking points within specific sections. The generation process considers narrative cohesion and the relevance of each talking point within its context.

\begin{figure*}[t]
\includegraphics[width=0.8\textwidth]{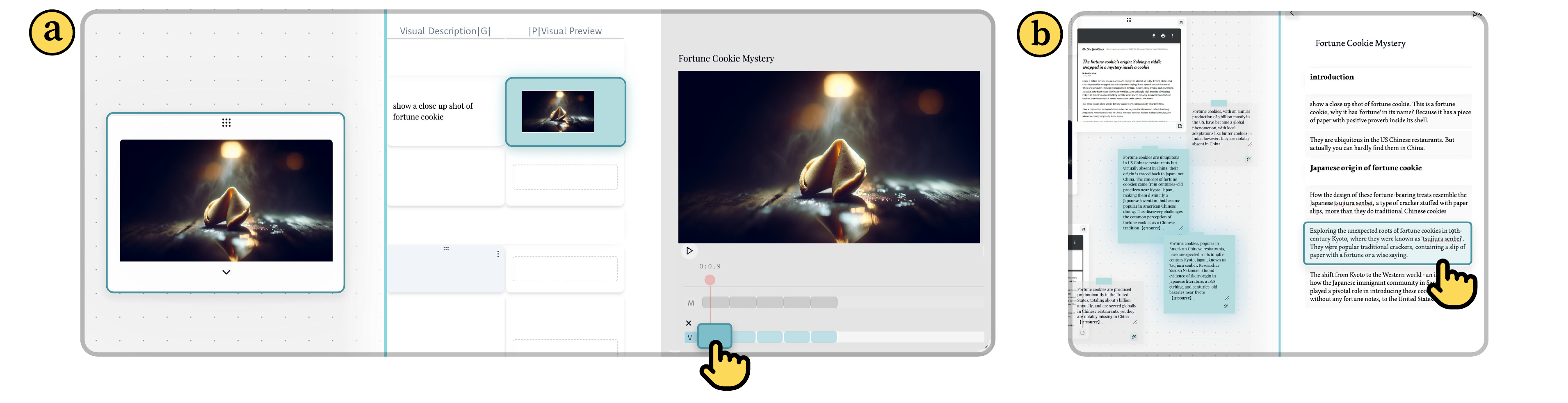}
\caption{Synchronized highlighting between different structures. (a) When the user clicks on a segment in the Timeline Editor, the corresponding Grid cell is highlighted, and the node is centered on Canvas. (b) hovers hover a talking point in the Narrative Editor, \system{} highlights relevant note nodes in the Canvas. }
\label{fig:synch-highlighting}
\Description[]{Two images labeled (a) and (b). Image (a) shows the \system{} user interface with a click icon hovering over a track segment in the Video Editor. (b) notes on the canvas are highlighted when the user hovers over a paragraph block in the narrative editor.}
\end{figure*}

\textbf{The Grid-Based Scene Planner} employs columns and rows to plan different elements in a scene chronologically.
 The grid-based scene planner also defines four types of columns: the \textit{Storyline} column houses the talking points; the \textit{Script} column contains the transcript for each scene; the \textit{Visual Description} column describes what visuals to show in each scene; and the \textit{Visual Preview} column shows the visual assets to be included in the video.  Besides basic operations such as adding, deleting, or shuffling rows and columns, users can populate each cell of the grid manually or generate content based on the context provided by existing rows and columns.

\textbf{The Timeline Editor} allows creators to preview the video and adjust its pacing. It organizes content into three types of tracks: audio, visual, and caption tracks. Each track comprises sequences of snippets as the smallest manipulable units in the timeline. While we did not implement AI features within the timeline, some techniques explored in prior work, such as aligning visual beats with audio ~\cite{davis2018visual} and transition suggestions ~\cite{veedio} can be incorporated.

\subsection{Cross-Structure Synchronization and Transformation with AI}
\label{sec:ux-across}
Informed by the desired interconnections (Section~\ref{sec:formal-interconnecton}),  we herein describe the cross-structure synchronization in \system{} in terms of the corresponding units, synchronization techniques, updating mechanisms, and the AI integration to support the synchronization. 

\subsubsection{ Canvas $\Leftrightarrow$ Other Structures (Assest Managment)} 
The canvas serves as a centralized asset hub, where any item added to other structures is automatically added as an asset node. To facilitate referencing relevant assets, we implemented \textit{bi-directional, automatic, synchronized highlighting}. Each canvas node corresponds to the smallest unit in other structures (i.e., a paragraph block in the narrative editor, a cell in the grid, or a snippet in the timeline). 
When a user edits a unit in another structure, related canvas nodes are dynamically highlighted (Fig. ~\ref{fig:synch-highlighting}). In this process, AI is incorporated to actively calculate relevance using embedding vectors of the content. Conversely, when a user hovers over a Canvas node, related units in other structures are highlighted. This ensures that all important assets are effectively incorporated into the video.

\begin{figure}[hb]
\centering
\includegraphics[width=0.85\linewidth]{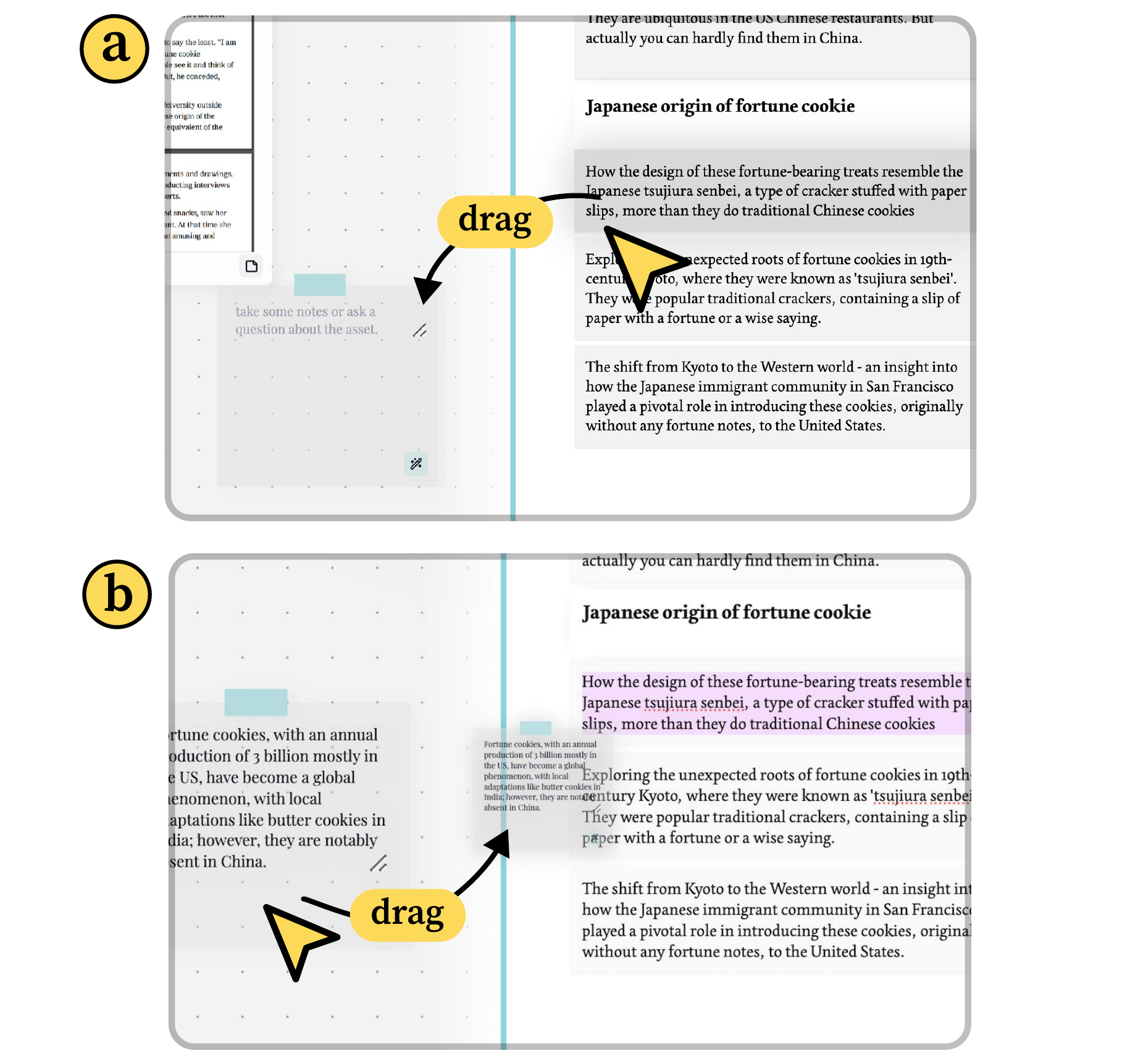}
\caption{Transformation of information between Canvas and text editor. The user can (a) drag a talking point into an empty note to extract relevant content from the linked article; or (b) drag a note into the talking point to revise or create a new talking point based on extracted information from the article.} 
\label{fig:Canvas-editor}
\Description[]{Two images (a), (b) all show the Canvas and Narrative Editor. In (a), a cursor icon is shown to drag a paragraph in the Editor onto a note in the Canvas. In (b), a cursor icon is shown to drag a note into the paragraph in the Editor.}
\end{figure}

\begin{figure*}[ht]
\includegraphics[width=1\textwidth]{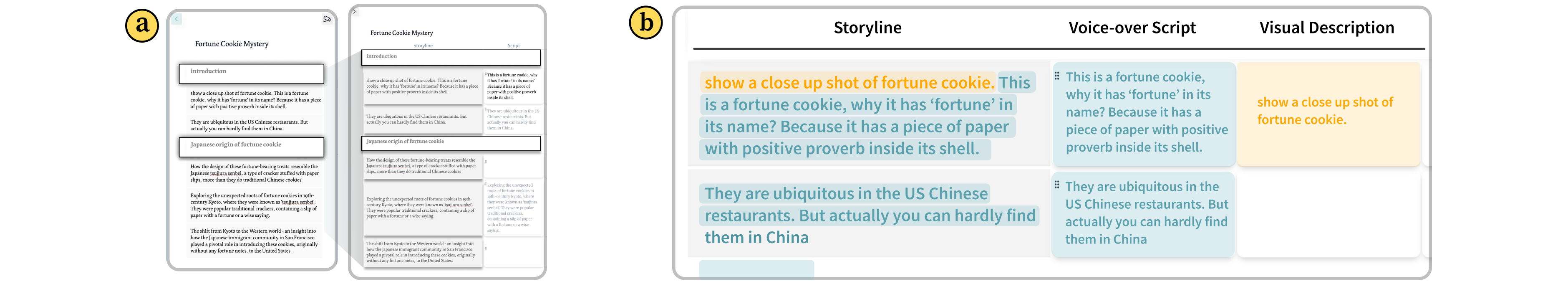}
\caption{Transformation of information between the Narrative Editor and Grid. (a) Content in the text editor transforms into a storyline column in the Grid. (b) Population of the narrative content into Grid columns based on different column types.}
\label{fig:editor-Grid}
\Description[]{Two images (a) and (b). Image (a) shows an arrow between Narrative Editor view to Storyline in Grid and click icon hovering over a button in the Editor. Image (b) shows the Grid with Storyline, Script, and Visual description columns. Script and Visual description have gray text within them.}
\end{figure*}

\begin{figure*}[ht]
\includegraphics[width=1\textwidth]{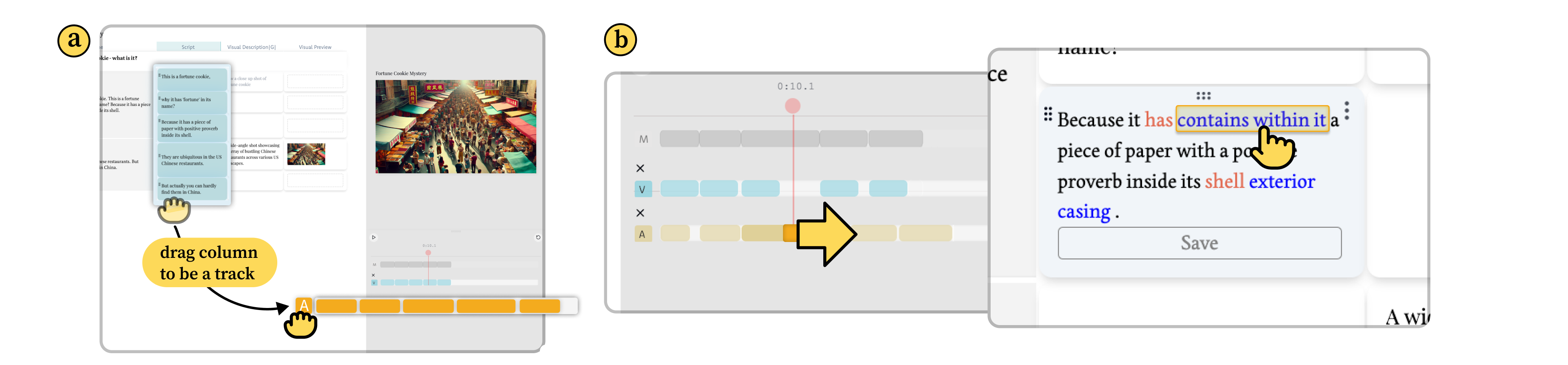}
\caption{Transformation of information between Scene Planner and Timeline Editor. (a) Drag the script column to transform it into an audio track. (b) Fine-tune the time segments and adjust the corresponding script.}
\label{fig:time-Grid}
\Description[]{Two images (a) and (b). Image (a) shows a hand icon dragging a script column in the Grid to the Video Editor. Image (b) shows a snippet of the tracks in the Video Editor and a script cell with black, red, and blue text with a `Save` button below. }
\end{figure*}

\subsubsection{Canvas $\Leftrightarrow$ Narrative Editor (Narrative Development)} 
To help users form cohesive narrative points from fragmented notes, we implemented \textit{bi-directional, user-initiated, synchronized editing} between nodes in the canvas and the paragraph blocks in the narrative editor. When a user drags a note node into the narrative editor, \system{} leverages generative AI to transform the note content based on the drop target: dropping into a new block transforms its content into a new talking point with the existing narrative sequence as context; dropping into an existing block revises the content to integrate the note while preserving the original meaning (Fig.~\ref{fig:Canvas-editor}b). Conversely, when users drag a talking point into an empty canvas note node, \system{} extracts content relevant to that point from the note's associated asset content to further support this process (Fig.~\ref{fig:Canvas-editor}a).

\subsubsection{Narrative Editor $\Rightarrow$ Scene Planner (Scene Development)} 
To provide an effective starting point for scene planning, we implemented \textit{uni-directional, user-controlled, synchronized editing} from the narrative editor to the grid-based planner. The rows in the grid correspond to the different types of blocks in the narrative editor: a \textit{section row} corresponds to a section block in the text editor (Fig.~\ref{fig:editor-Grid}a); a \textit{talking point row} corresponds to a paragraph block. 
When users modify content in the narrative editor, ~\system {}  utilizes LLMs to categorize the content based on the grid’s columns (e.g., visual descriptions or voiceover scripts) and maps it to the corresponding grid cells as suggested content (Fig.~\ref{fig:editor-Grid}b). Users can press ``Tab" to quickly accept the suggestion or overwrite them.

\subsubsection{Scene Planner $\Leftrightarrow$ Timeline Editor 
 (Temporal Adjustment)} 
To support both structural and fine-grain temporal adjustments, we implemented \textit{bi-directional, synchronized editing} between the grid-based scene planner and timeline editor. The timeline can be seen as a transposed view of the scene planner, with each column corresponding to a track and individual grid cells representing specific time segments within those tracks. Users can drag columns to the timeline to form default tracks (e.g., the script column becomes an audio track, and the visual preview column becomes a visual track) (Fig. ~\ref{fig:time-Grid}a). AI facilitates the transformation between grid cells and their corresponding time segments, with different updating mechanisms for each direction. Grid-to-timeline updates are \textit{automatic}. For example, extending a script or replacing a video clip will automatically adjust the corresponding time segments. Timeline-to-grid updates are \textit{user-controlled}. For example, when a user shortens a time segment, \system{} suggests edits to the corresponding grid cell to align with the new timing while preserving the original meaning. Users can accept these suggestions or make their own edits (Fig. ~\ref{fig:time-Grid}b).

\subsection{System Walk-through}
\label{sec:ux-walkthrough}

\begin{figure*}[!ht]
\centering
\includegraphics[width=1\textwidth]{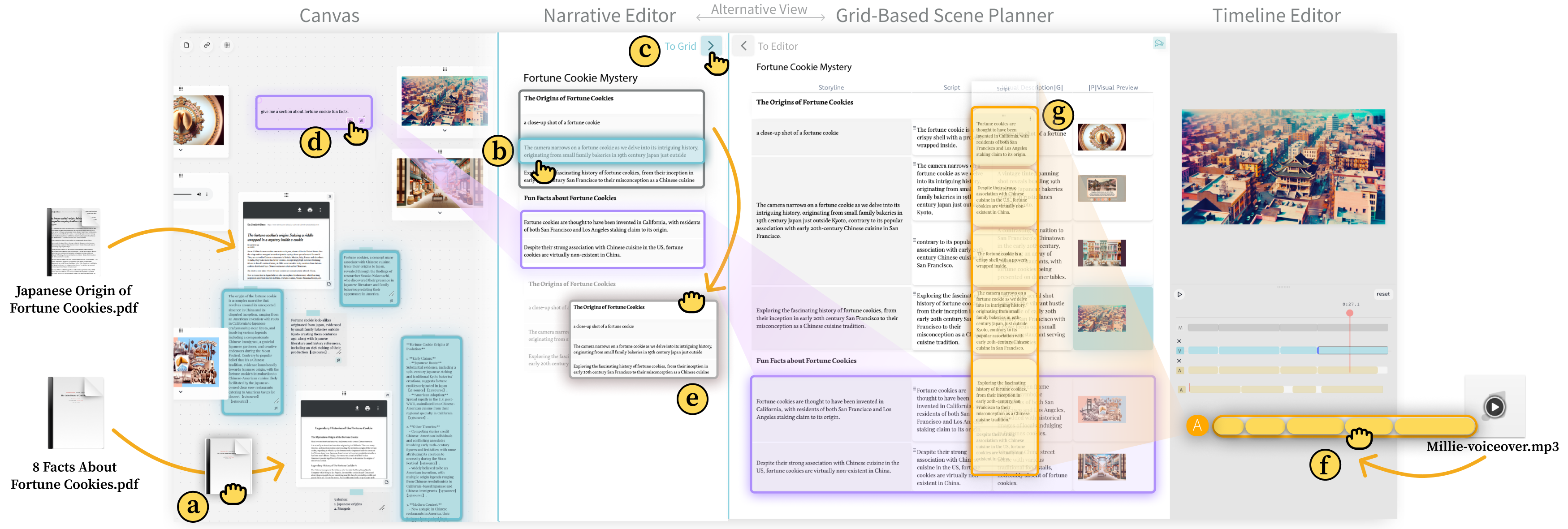}
\caption{A user planning and creating a video using \system{}. The user can import external assets to the canvas (a); while the user is writing in the Narrative Editor, relevant notes are highlighted (b); the Narrative Editor expands and transforms into the Grid-Based Scene planner (c); the Prompt Node enables the user to generate a section of the narrative automatically (d); the user can re-arrange sections of the narrative by dragging and dropping in the Narrative Editor (e); the Timeline Editor enables users to import audio files (f) and automatically align the imported audio with the existing script (g).}
\label{fig:walkthrough3}
\Description[]{Multiple interactions are shown on the \system{} interface. The structures and interactions are labeled a-g. Interaction (a) the Canvas with external PDFs and images added onto it. (b) shows the Narrative editor with sections and paragraph blocks. (c) shows a paragraph being clicked on and relevant notes being highlighted. (d) shows the Grid Based Scene Planner with the script, visual preview and, visual columns filled out. (e) shows the Timeline editor. (f) shows a click on a button from a sticky note that says ``generate the fun fact section'' with arrows pointing to the generated paragraphs in the Narrative Editor and the generated rows Scene Planner Grid, (g) shows a drag action from an external audio file to a track in the Timeline Editor with an arrow to the aligned script column in the Scene Planning Grid.}
\end{figure*}

We walk through a scenario where a creator, Millie, uses \system{} to make an explainer video about ``The Mystery of Fortune Cookies''. Millie first imports some assets she has collected about fortune cookies into \system{}'s canvas (Fig.~\ref{fig:walkthrough3}a). 
Using the embedded AI functionality in the canvas, she explores the materials and conceptualizes two high-level sections for the video: \textit{``The Origins''} and \textit{``Fun Facts''} and adds them to the narrative editor.

\subsubsection{Bottom-Up Creation} Millie wants to open the video with a familiar image, so she writes ``\textit{a close-up shot of a fortune cookie}''  as the opening sentence. 
When transitioning to its Japanese origins, she recalls an article but forgets some details. \system{} highlights relevant notes in the canvas as reference (Fig.~\ref{fig:walkthrough3}b). To seamlessly incorporate the facts, Millie drags a note into the paragraph, transforming it into a talking point while ensuring a cohesive narrative flow that integrates the new information (Fig.~\ref{fig:Canvas-editor}b).
Satisfied with the story, Millie moves to the scene planner by clicking on the toggle located on the top right corner of the editor (Fig.~\ref{fig:walkthrough3}c). \system{} transforms the content in the narrative editor to automatically populate the scene planner (Fig.~\ref{fig:editor-Grid}), giving Millie an effortless first draft. With the synchronization, she also gets a rough cut in the timeline editor with a generated voiceover based on the script. Millie adds visuals, previews the video, and refines the scenes by iterating between the scene planner and the timeline editor.

\subsubsection{Top-Down Creation} For the second section, Millie starts by prompting \system{} to generate the entire section first (Fig.~\ref{fig:walkthrough3}d). Instead of directly presenting her a video cut, \system{} follows the structures and populates talking points, voiceovers, visuals, and time segments within each structure progressively. This allows Millie to review and refine as the video takes shape. She can reorganize sequences in the narrative editor to explore different narrative flows (Fig.~\ref{fig:walkthrough3}e).
As changes propagate across different structures, she can easily make further edits or preview the new version. Finally, Millie drags her own voiceover recording into the timeline to replace the generated audio (Fig.~\ref{fig:walkthrough3}f). \system{} stores it as an asset in the canvas, as well as intelligently aligns her audio with the existing script in the grid, and adjusts the timing of each scene (Fig.~\ref{fig:walkthrough3}g).

With the help of \system{}, Millie successfully creates her first cut of the explainer video. \system{} yields not only the output video, but also the artifacts of the process: the assets and notes on the canvas, the narrative in the editor, and the scene plan in the grid. These artifacts assist Millie in verifying sources and making future edits.

\subsection{Implementation Details}
\system{} is built using TypeScript with React for the front end, Zustand for state management across different views, MongoDB for the database, and a Python back-end. AI features include OpenAI's GPT-4 API for text generation, DALL·E 3 for image generation, and Whisper 1 for converting text to voiceover. 
A detailed description of the implementation is included in Appendix~\ref{sec:apx-prompt}.

%% file: tex/06_evaluation.tex
\section{Evaluation}
\label{sec:eval}
We evaluated \system{} with a user study to investigate whether the design approach has resulted in an effective co-creation environment by answering the 
following research questions:
\begin{itemize}[leftmargin=2.5em]
\item [\textbf{RQ1}] Whether the aggregated compositional structures can serve as an effective common ground for human-AI collaboration;
\item [\textbf{RQ2}] Whether AI can reduce challenges associated with completing and synchronizing the compositional structures;
\item [\textbf{RQ3}] Whether there are new workflows and usage patterns enabled from this design approach;
\item [\textbf{RQ4}] What tensions may arise in the co-creation environment?
\end{itemize}

\subsection{Study Procedure}
We evaluated \system{} with different creator profiles (i.e., experts and novices) and different creation workflows (i.e., human creates and AI synchronizes as well as AI creates and human refines) to more broadly evaluate the environment's effectiveness. The inherent complexities of video creation, coupled with the learning curve associated with a new authoring interface and its various features, resulted in the entire study exceeding 3 hours. Therefore, we broke the study into two parts. In the first part, all participants were asked to create a video with a bottom-up approach (i.e., human creates and AI synchronizes). In the second part, participants started with a single prompt to create a video, after which they evaluated and iterated upon it to achieve their desired outcome.

We recruited 6 novices (N1-N6, 3 female, 3 male), and 4 experts (E1-E4, 3 female, 1 male) for the study. Novices self-reported having limited knowledge of video creation tools and techniques and have created less than ten videos. All experts had at least two years of video editing experience and published videos either monthly (E4) or once every few months (E1-3).  All participants attended Part 1 of the study. Novices and experts were compensated with \$40 and \$100, respectively, for their participation in Part 1 (2 hours). Three novices (N1-3) and three experts (E1-3) attended Part 2 (1.5 hours), and received \$25 and \$50, respectively.

\subsubsection{Part 1: Bottom-Up Creation} In this 120-minute study, participants (6 novices, 4 experts) were tasked to create a 30-second video about fortune cookie origins from scratch using our system. Before the study, participants were given two articles to familiarize themselves with the video topic. Seven studies were conducted in-person and three remotely via Zoom. All participants accessed \system{} through a web browser.

\hspace*{1mm}\emph{Introduction and System Walk-through ($\sim$40 minutes)}. 
The experimenter first introduced participants to \system{} and interviewed them about their experience with video creation and generative AI. Next, the experimenter walked the participants
through \system{}’s features by guiding them in creating the first half of the video (introduction to fortune cookies). During the walkthrough, 
the experimenter explained each interaction and asked participants to perform specific actions. 

\hspace*{2mm}\emph{Creation task ($\sim$40 minutes)}. 
Participants were asked to complete the second half of the video (focusing on the cookie's Japanese origins). The system was pre-loaded with assets they could optionally use during their creation process. 

\hspace*{2mm}\emph{Iteration task ($\sim$15 minutes)}. 
Participants were given an additional article describing stories of fortune cookies. They were instructed to extract narrative points from the article and revise the video to add at least one narrative point.

\hspace*{2mm}\emph{Questionnaire and Post Interview ($\sim$25 minutes)}. 
After completing all the tasks, participants filled out a questionnaire
about the usefulness of \system{}'s concept, features, and their experience, followed by a semi-structured interview to gather further insights.

\subsubsection{Part 2: Top-Down Creation} In this study, participants created another 30-second video. To allow the participants to easily evaluate AI-generated content (e.g., narrative, images, video sequences), we asked the participants to prepare materials they were familiar with, such as a novel, article, or blog post, as the information source. Three studies were conducted in-person and 3 via Zoom.

\hspace*{2mm}\emph{Task Introduction and System Revisit ($\sim$10 minutes)}. 
Participants were first given a general introduction to the task, followed by a quick walk-through of the system to serve as a refresher.

\hspace*{2mm}\emph{Practice Task ($\sim$10 minutes)}. 
Participants were instructed to use AI to generate a video with the content they provided by writing a simple prompt such as `create a video'. The goal of this step was to help them get a sense of the generation process and the generated video.

\hspace*{2mm}\emph{Creation Task ($\sim$45 minutes)}. 
Participants were asked to write more detailed prompts to generate a video and then refine the generated video through iterative adjustments using the system.

\hspace*{2mm}\emph{Questionnaire and Post-study Interview ($\sim$20 minutes)}. 
After completing all the tasks, participants completed a questionnaire
about the usefulness of structures in iteration and comprehension, followed by a semi-structured interview to gather further insights and compare their experience with the bottom-up approach.

\section{Findings}
Results from the questionnaires and interviews provide evidence of using compositional structures to ground human-AI collaboration and facilitate fluid iteration of various aspects of the video. The compositional structures and generative AI together yield a new cost structure ~\cite{coststructuresensemaking} for video creation, enabling new workflows and unveiling exciting research questions that require further investigation. We have included a few samples of the outputs created during our user study in the Appendix ~\ref{sec:output}, and more results created during our user study can be found in this gallery.\footnote{\system{} User Study Results: \href{https://videorigami-userstudy.netlify.app/}{https://videorigami-userstudy.netlify.app/}}

In the sections that follow, we describe our findings in terms of the four research questions and the general implications we draw from these findings that can potentially apply to other creation domains in the use of compositional structures for human-AI co-creation environments.

\subsection{Compositional Structures as Strong Foundation for Collaboration (RQ1)}
\subsubsection{Effective Representation of the Whole Picture} By aggregating multiple compositional structures inherent in video creation together, \system{} preserves and visualizes the entire creation context and progress. Experts found the structures familiar, and novices found the system instructive and could help them develop an understanding of video creation and ``easily get hands-on'' (N3) using each structure to compose different aspects of a video. 

\begin{quote}
The way you are putting those things here has a structured layout which helps me to understand how the video is composed. (N3)
\end{quote}

Additionally, both experts and novices reported that the system enabled them to stay aware and oriented in a typically messy process. They could ``understand what's going on'' (E3) and ``pick up wherever they want'' (N5). Interestingly, while we were targeting human-AI collaboration, multiple participants commented on the system's suitability for human-human collaborations, indicating \system{}'s effectiveness as a general collaboration platform.

\begin{quote}
I like that the context is always in front of me so I'm not blinded by anything. (N2)
\end{quote}
\subsubsection{Facilitating Understanding Generated Content and Generation Capabilities} Participants found these structures particularly effective in helping them comprehend AI-generated content and identify issues in AI generation. E2 mentioned that the structures allowed them to ``directly make sense of the generated results''. E1 mentioned that by reading the generated section headings in the narrative editor, they could \textit{``quickly know how AI suggests the story should go''}. N2 noted on the grid structure in the scene editor \textit{``gives me more transparency about what goes into each frame''}. 

\begin{quote}
I like the fact that I know what the breakdown is. I'm seeing what the script is, and I'm already thinking about what the visuals seem to look like. And then I can quickly spot something that does not match my expectations and know where I should change. (N2)
\end{quote}
Beyond helping creators understand what AI is generating, structures also facilitate the understanding of what AI can generate, offering creators a glimpse of achievable results. 
\begin{quote}
I see the Rose and the Bob in the image, but that has nothing to do with the text in this area. So it tells me that there's some global understanding of things here... that makes me feel confident about realizing there is some cohesion to this, and I can maybe, through better prompting, get somewhere. (N1)
\end{quote}
The ease of staying aware of the whole picture of video creation and comprehending AI's work and capabilities as a collaborator bolstered participants' awareness and assurance 
in collaborating with AI on the video creation task. 

\subsection{Leveraging Generative AI to Complete and Synchronize Compositional Structures (RQ2)}
All participants reported that the synchronization (Mean=5, SD=0) and generative function (expert M=5, SD=0; novice M=4.83, SD=0.41) increased their productivity by helping them quickly get to a rough cut from a blank canvas.

\subsubsection{Shortening the Path to the First Rough Cut}

As mentioned in the formative study with video creators, a unique challenge of video creation is the lengthy process required to reach the first rough-cut preview compared to other tasks such as writing or graphical design. Creators must constantly speculate about the final audio-visual outcome while collecting assets, developing narratives, and planning the scenes until the very last stage. Yet, they are frequently surprised by the significant gap between their expectations and the actual outcome. However, the time constraint at the last stage of the workflow leaves little room for major structural changes. A significant benefit of the unprecedented speed enabled by AI is the ability to quickly create a rough cut to examine whether the final result matches their expectation. This was found particularly useful by participants, as it allowed them to quickly get a sense of whether the narrative is effective.

\begin{quote}
I think it was a good start because usually, I feel the most difficult part is to get a timeline out. With a timeline, it will be much easier to tweak. (E2)
\end{quote}

\begin{quote}
I'm struggling to even build that initial structure; that's when I would like to have this as my starting point and then build on it. (N2)
\end{quote}

\subsubsection{Parallel Development and Iteration of Multiple Compositional Structures}
Different from what we found in the formative study, where creators tend to complete one structure as much as possible before they move to the next, during both parts of our study, we observed a frequent switch across the structures throughout the creation process. At the initial stage, N1 started with exploring the canvas, while N5 started by generating multiple talking points in the narrative editor and brainstormed from there. N2, and E3, on the other hand, switched frequently between the narrative editor and scene editor for narrative development. After they had parts of the story, some creators (N4, E4) previewed the videos frequently when making iterative edits, while other creators (N3, E1) frequently switched between the scene editor and canvas. This usage pattern proved creators' desire for and \system{}'s ability to support inspecting and manipulating different compositional structures during creation. 

All creators strongly agreed that the synchronizations made their creation processes more productive by reducing the effort of \textit{``moving things around''} (E2). Some commented that this also made them more creative by allowing them to focus on the creative aspect of the process (N5, E1, E2). 

\subsubsection{AI that Went Unnoticed} Because the synchronization among the compositional structures follows the constraints of the structures, they are less error-prone than generating an entire narrative from talking points or generating the visuals based on text descriptions. As a result, the intelligent synchronization often went unnoticed. In the current system, synchronization is approached conservatively, where we use visualizations to show connections and update the content in a suggestive manner.  However, feedback from participants indicates a desire for more proactive synchronization. For example, when switching between narrative editor and scene planner, multiple creators expressed preferences for automatically transforming their storyline into a revised version of the voice-overs or visual descriptions. 

\begin{quote}
The live synchronization, it just seemed so natural like that's how it should be, I did not realize I was using it all the time. (N1)
\end{quote}

\begin{quote}
    Since I need to review it anyways, I would prefer to have it [AI] directly transform my wording.  (N3)
\end{quote}

\subsection{Emergent Workflows Due to Shifts in the Cost Structures (RQ3)}
In addition to the parallel development and iteration of multiple compositional structures, we observed other novel workflows due to the shifts in the cost of content creation and evaluation process.

\subsubsection{Grid-Based Scene Structure as the Main Playground}

Despite the different usage patterns across participants, the grid-based scene planner stood out as the pivotal structure for the entire creation process when using \system{}. E1 described it as their \textit{``main battlefield''}. This is different from what we found in the formative study, where creators usually spend most of their time on the narrative and timeline, as they did not want to be constrained by a rigid grid structure and considered completing a detailed scene planning in the grid too much work. 

With \system{}, the grid-based scene structure not only organizes and displays all the materials used in the video but also connects with the narrative editor and the timeline. Participants realized they could command all the materials, and any changes made in the scene planner could propagate to the narrative structure and timeline. As a result, they felt comfortable entering the grid much earlier in the creation workflow and sticking with it. 

\subsubsection{Hidden Cost of Generation} 
With significantly lowered costs of generating and synchronizing the structures, we expected participants to leverage these capabilities to explore different video ideas and perform more large-scale revisions. However, these happened less than what we expected. This could be because of the constraints inherent in the study settings, such as the limited time participants had to explore different ideas and their limited commitment to the outcome. Nevertheless, we also observed evidence of how the low cost of the interaction and automation techniques disguise other costs in the entire creative activity.

In the first part of the study, participants were requested to develop the video piece by piece. Because of the low cost of using AI to generate content, we observed participants using AI to quickly fill the narrative and scene structure without giving too much thought. In particular, participants reported that filling in the empty cells in the grid was very tempting, as they could quickly \textit{``have something look good and get a story across''}(N1). As a result, while some of the generated images and text were not ideal, they often opted for filling the entire structure instead of spending effort refining specific elements. Yet, the cost of structural revision of the video quickly rises as the video becomes more complete, discouraging participants from large-scale iterations.

In the second part of the study, given the low cost of generating a video, we expected participants to explore different ideas by prompting several rounds. However, only 2 participants (N3, E3) made section-level changes to the generated storyline. We observed several reasons. 
First, participants often found the generated content good enough as a cohesive piece, albeit different from what they expected (E1, E3). More importantly, while the cost of generating a video was significantly lower, the cost of assessing a video remained high, including the time spent waiting for the video generation, reviewing the video, comprehending the narrative, and reviewing the visual prompts. As a result, participants were incentivized to accept a good enough generation and only perform small iterations.

\begin{quote}
Now that I see it. I feel like alien invasion is a good starting point because it captures people's attention. Ok, I actually think the first part [of AI-generated narrative] makes sense to me now. (E1)
\end{quote}

\begin{quote}
I was mainly looking at how the storyline was generated and compared it to what I was expecting… It's interesting that it only did the beginning of my prompt, but I like the storyline it gave. It makes sense. So I'm not going to tweak anything with the storyline, mainly the script and visual description. (E3) 
\end{quote}

These findings suggest that while the combination of compositional structures and generative AI reduces the cost of manual operation, it does not magically reduce all costs but may also introduce new ones. More specifically, our findings indicate that lower operation-wise costs do not necessarily lead to increased iterations. The high-fidelity designs that AI generates create the impression of completeness, which may discourage further iterations.

\subsection{Tensions of Creative Expressions (RQ4)} 
While all participants mentioned the generative functions for filling the structures made them feel productive, their opinions on whether this made them creative were mixed. Some participants appreciated that the generated results could provide them with ideas they had not thought of, hence \textit{``enlarged creativity realm'' }(E1). Others found the generated results impeded their creativity, despite knowing they could create everything manually. Novices and experts also differ in their perceptions. With a 5-point Likert scale, novices (M=4.5, SD=0.84) reported feeling more creative when using the generative functions compared to experts (M=3.5, SD=0.96); novices (M=4.33, SD=0.52) also felt a greater sense of transparency and control over the AI functionalities compared to experts (M=3.5, SD=1).

\subsubsection{Overshoot, Undershoot, and Sweetspot of Generation Fidelity} Some participants found the high-fidelity content generated by AI overwhelming and impeded their creative thinking. \textit{``Sometimes, I have a vague idea in mind, but the generated visual is so detailed that, upon seeing it, I find my thoughts blocked''} (E4). This indicates that the concrete nature of generated visuals can overshadow nascent ideas. On the other hand, creators felt frustrated and a sense of \textit{``lost control''} when AI failed to generate their envisioned visuals. Interestingly, the visual description generated by AI, which described the intended visual for a scene,  was positively received by all participants. We observed both novices and experts, regardless of whether the visuals matched their expectations, frequently inspect AI-generated descriptions of the visual content, as they found AI's description of suitable visuals, compared to actual visual content, both informative enough and leaving room for their own ideas. 

\subsubsection{Creative Expression with Prompts} We observed that participants' expertise in prompting significantly impacted their creation experience. Participants who lacked prompting experience often got confused about why AI produced certain results, whereas participants who are experienced with generative AI tools could better comprehend issues in the generated results. 
\begin{quote}
By comparing the visual descriptions and the images in these rows, I know it got confused by this abbreviation. (N3, comprehending AI's mistakes)
\end{quote}

\begin{quote}
Originally, I thought that the script would be the poem, line by line, right? But I don't know where this is coming from. It's like a third-person perspective. Just wonder what's the logic here.. (E1, confused by the AI-generated storyline)
\end{quote}

Participants' abilities to comprehend AI-generated content and to express their desired outcomes in prompts directly affected the effectiveness of their iterations. Being skillful at prompting enhanced participants' sense of control during the collaboration, broadened their horizon of what content was achievable, and maintained their creative engagement with the creation process.

\subsubsection{Contrast between Novices and Experts } The contrasts between novices' and experts' perceptions towards AI-generated content were also observed in our study.  Experts, such as E2, reported concerns about the decline in their engagement with the creative process.  Novices, on the other hand, were much more receptive to AI-generated content, as AI empowered them to create content beyond their abilities. 

\begin{quote}
It’s so fast to generate results, and the result makes sense. I found myself stopping thinking during the process... If I am emotionally invested in a project, I might be hesitant to use AI, so I won't lose commitment to it. (E2)
\end{quote}

\begin{quote}
Because I know that I lack the ability to do a lot of things ... so this (AI) really gets me going. (N1)
\end{quote}

As discussed in the previous section, participants' expertise in prompting AI significantly affected their creation outcomes and experiences. Within the research team, we found videos created by novices with significantly more prompting experiences exhibited higher quality than those created by expert participants during our study. It is imperative to note, however, that this observation does not serve as evidence suggesting that novices with proficiency in prompting are capable of creating content of a higher caliber than experts. Experts, by virtue of their extensive knowledge and understanding of the principles underlying the production of high-quality videos, maintain a distinct advantage. Nonetheless, the findings of our study suggest that expertise in prompt engineering can, to a certain degree, reduce the gap between novices and experts in creating creative content.

\subsection{Summary}
Findings from the user evaluation of \system{} provide evidence of the effectiveness of the design approach of employing compositional structures as the foundation of the human-AI co-creation environment. Specifically, we found the aggregated and interconnected compositional structures enabled creators to stay oriented throughout the creation process and facilitated their understanding and control of AI generation. Participants acknowledged that they felt more productive with AI helping them complete and synchronize the structures. We also observed new workflows and costs associated with human-AI video co-creation, which we discuss further below.  

%% file: tex/07_discussion.tex
\section{Discussion and Future Work}
\label{discussion}
The user evaluation allowed us to understand the effectiveness of the human-AI video co-creation environment created with our design approach. We first discuss how the cost structure of the video creation activity is shifted and then discuss the implication of the design approach by situating it in the broad scope of information work and activity-centered information spaces. 

\subsection{New Costs in Human-AI Video Co-Creation and Beyond }
Besides the commonly recognized challenge of prompting AI to generate the desired content, our study uncovered three new types of costs. While these costs were observed in the context of video creation, they are broadly applicable to other domains within human-AI co-creation.

\subsubsection{Evaluation Cost Inherent in Content} The high cost of evaluating a generated video became prominent in our study, which made participants reluctant to explore different video ideas. Participants not only had to examine the final generated content but also the underlying narrative and the reasons for AI generating the visuals for specific scenes. Among the various types of AI-generated content, images only require a glance to determine their suitability, whereas other types of content, such as text and video, require significant cognitive effort to consume and evaluate. To reduce the evaluation cost of text, research has explored transferring text to diagrams or summarizing the text to facilitate consumption of a large amount of text \cite{graphologue2023, textsummarizationreview}. 

Future research should investigate how to apply these ideas to reduce the evaluation cost of video and other content emphasizing the congruent, narrative, and temporal aspects, as these aspects inherently demand significant cognitive loads to evaluate.

\subsubsection{Harms of High-fidelity Content} AI-generated content is often of high fidelity, which can be harmful to early-stage design, as the high-fidelity content pushed the participants toward early convergence when they could benefit from more diverse ideas \cite{Prototypingdynamics, parallelprototyping, righdesign, detailinstoryboard, suh2024luminate}. One approach to mitigate this problem is to present users with multiple options, which was found to encourage exploration in design \cite{parallelprototyping}. However, as mentioned above, evaluating a generated video incurs significant cost, let alone reviewing and comparing multiple options. Another option is instructing AI to purposefully generate low-fidelity prototypes to avoid nudging users to converge too early or generating design space to encourage exploration \cite{suh2024luminate}. 

Future research should explore combining these approaches while developing new strategies to mitigate the potential harm the high-fidelity content may pose to users’ agency and creativity.

\subsubsection{Fear of Losing Previous Versions} Popular AI tools such as, ChatGPT and Midjourney, utilize chat as the primary interaction mechanism. A benefit of the chat mechanism is that all previous prompts and results are automatically preserved, enabling users to recover to a previous version easily. \system{} avoided using chat due to natural language's inefficiency in supporting flexible reference and manipulation of inherently spatial structures \cite{TenMyths}, and instead employed the direct manipulation paradigm \cite{hutchins1985direct}. The downside, however, is that every edit operation is destructive. While undo/redo and version control can be provided, they typically do not make history as directly visible and easily retrievable as the chat mechanism. We found this contributed to participants' reluctance to iterative prompt engineering, which is often seen in other AI-assisted workflows. Future work will explore how to incorporate the version control mechanism to meet the same level of visibility and accessibility.

\subsection{Generalizability of the Design Approach}
Synthesized from the literature across diverse domains, our design approach—identifying, designing, synchronizing, and integrating compositional structures—provides a robust framework for developing human-AI co-creation environments and is inherently generalizable to the domains we surveyed, including writing, music, podcast, and interactive media. We believe the design approach can also generalize to other types of content of a compositional nature, such as game and VR scene development. We discuss key design considerations when applying the design approach.

\subsubsection{Devising New Compositional Structures}
 Our design approach emphasizes identifying compositional structures within existing workflows. However, we recognize that novel compositional structures can often bring significant benefits to the content creation workflow ~\cite{DataInk, Stickylines, underscore, truong2016quickcut}. For example, time-aligned transcript \cite{semanticSpeechEditing} has significantly reduced the editing effort for audio and video content \cite{truong2016quickcut, crosspower, contentbased} by enabling creators to more easily attend to the temporal and congruent aspects of the content. When designing co-creation spaces, it may be fruitful to consider what content aspects receive inadequate support and devise new compositional structures to fill the gap.

 \subsubsection{Supporting Freeform Exploration and End-user Customizable Structures} Freeform canvas is a common structure used in all domains we surveyed, despite that it does not enforce specific compositional rules. This is because freeform canvas enables creators to experiment freely with different compositions, especially during exploration. Structures that can be freely created and customized based on end-users' needs can also be beneficial. This may require the development of lower-level primitives with meta-compositional rules that describe how compositional structures themselves should be composed.

\subsubsection{Managing Multiple Compositional Structures within Unified Workspaces} An interface design challenge is how to effectively organize and integrate multiple compositional structures within the interface without incurring significant navigation and interaction costs. VideOrigami, for example, supports the merging of narrative editor and scene planner to reduce screen clutter and management costs. Domain-specific solutions may need to be devised when developing the co-creation space for other domains.

\subsubsection{Understanding the Shifts of Cost with Compositional Structures}
Our study revealed that creators' reliance on the structures may shift as they are integrated. For example, in our formative study, some creators skipped using the grid-based structures due to the high cost of developing and maintaining the structures. However, participants in our user study found the grid-based structure pivotal as it allowed them to engage with multiple content aspects. Therefore, designers of co-creation spaces need to understand the shifts of cost associated with integrating and synchronizing the compositional structures and their implications on users' workflows, and design the interface accordingly.

\subsection{Beyond Content Creation and Compositional Structures}
A clear next step of this work is to apply the design approach to other domains to further verify its generalizability. Beyond that, compositional structures are not the only structures employed in content creation. Structures that enable exploration of the design space, comparison of options, and tracking of changes are equally important in content creation workflows. Additionally, tasks like planning and decision-making often rely on various structures—for example, using tables to compare options during travel planning or data visualizations to facilitate filtering and ranking. A promising direction for future research is to develop a comprehensive taxonomy of these structures, along with methods for interconnecting them to allow seamless transformation of information across different structures. 

The human-AI co-creation environments are \textit{de facto} activity-centered information spaces that encompass various information structures and functionality traditionally distributed across individual applications \cite{informationspace}. Developing activity-centered, instead of application-centered, spaces has been a long-lasting endeavor in HCI \cite{bodker2021through, reframingdesktop, haijunRedesignSpace}. While previous attempts at this vision have failed for various reasons, we believe pursuing human-AI co-creation environments is another promising attempt. Based on the taxonomy mentioned above, we will develop a structure library that enables developers to easily create such environments or enable AI to intelligently compose such environments based on users' tasks.

%% file: tex/08_conclusion.tex
\section{Conclusion}
\label{conclusion}

AI is transforming not only how information is generated but also the fundamental structure of the information environment. In this work, we present an initial exploration of a design approach for developing human-AI collaborative environments. Specifically, we propose integrating compositional structures of information activities and embedding AI within and across these structures to create a cohesive, intelligent collaborative environment. Our findings from a video co-creation environment developed using this approach demonstrate that such an environment helps creators remain oriented within the creation workflow, gain greater control and interpretability of AI generation, and flexibly interweave human-driven and AI-driven processes. Grounded in the compositional nature of complex information content, with video creation as a representative activity, we believe this approach has significant potential for broad application across various domains.

%% file: main.bbl

%% file: tex/09_appendix.tex
\clearpage
\onecolumn

\section{Appendix}
\label{appendix}

\subsection{User Study Creation Outputs} 
\label{sec:output}
\begin{figure}[!ht]
    \centering
    \includegraphics[width=6in,keepaspectratio]{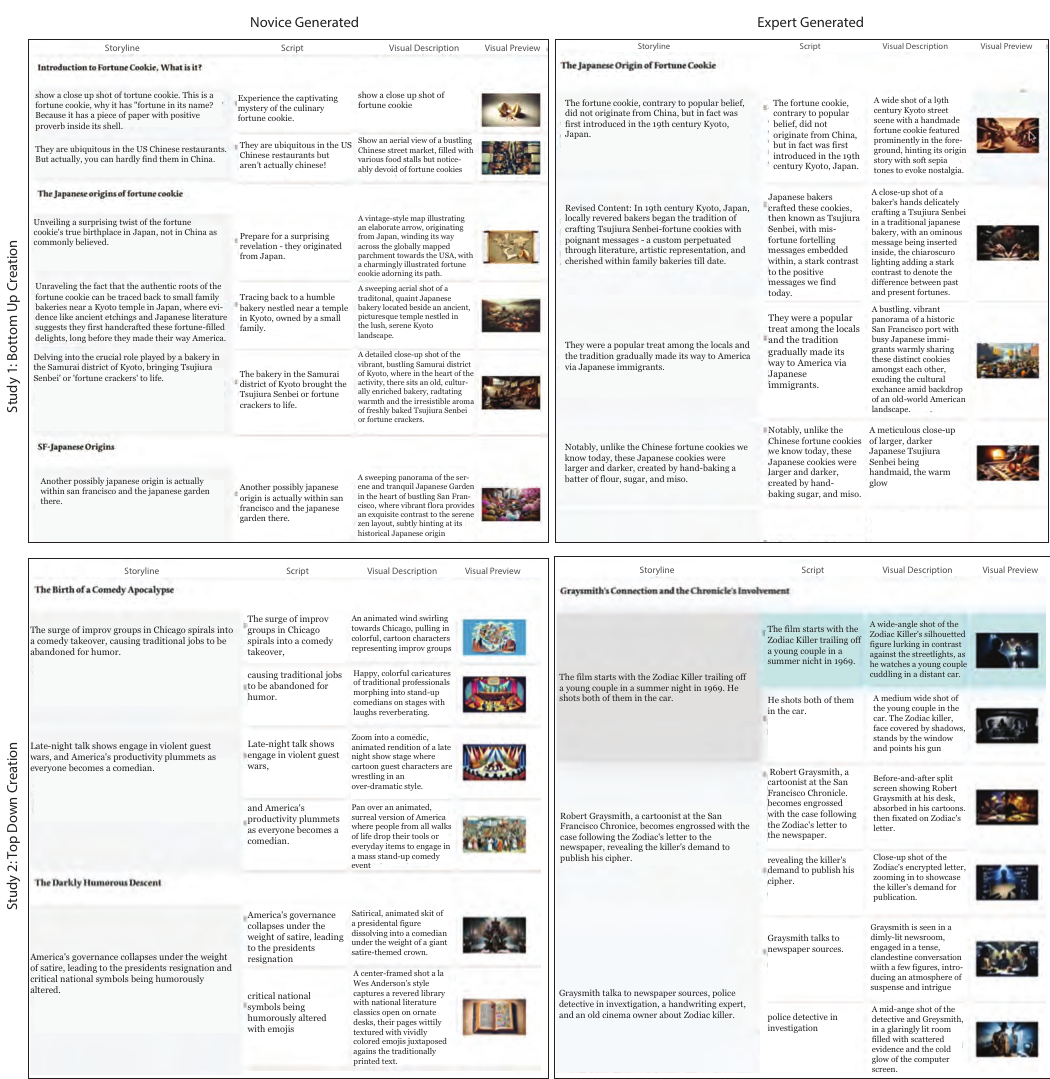}

    \captionof{figure}{Samples of outputs created during our user study. The top two outputs were created in Study 1. The bottom two outputs were created during study 2. Outputs in the left column were generated by novices while outputs in the right column were generated by experts.}
    \label{fig:results-figure}
    \Description[]{Four images total, two images in two rows. Each image shows a filled out Grid from \system{}'s interface. First row first column shows three sections, title `Introduction to Fortune Cookie, What is it?`, `The Japanese Origins of fortune cookie`, and `SF-Japanese Origins`. First row second column shows one section titled `The Japanese Origin of Fortune Cookie`. Second row first column shows two sections titled `The Birth of a Comedy Apocalypse` and `The Darkly Humorous Descent`. Second row second column shows a section titled `Graysmith's Connection and the Chronicle's Involvement`.}
\end{figure}


\newpage
\subsection{User Study Supplemental Data}
\label{sec:apx-survey}

\newcolumntype{Y}{>{\raggedright\arraybackslash}X}

\begin{table}[ht]
\centering
\label{table:user_demographics_table}
\small
\caption{Expert Participants Demographic Data.} 
\begin{tabularx}{\textwidth}{>{\centering\arraybackslash}m{1.8cm}|>{\centering\arraybackslash}m{2.3cm}|>{\centering\arraybackslash}m{2.5cm}|>
{\centering\arraybackslash}m{2.5cm}|>{\centering\arraybackslash}m{2.6cm}|>{\centering\arraybackslash}m{1.8cm}|>{\centering\arraybackslash}m{1.8cm}}
\toprule
\hline
\textbf{Participant} & \textbf{Video Creation Experience} & \textbf{Video Publishing Frequency} & \textbf{Video Creation Domain} & \textbf{Experienced with AI Video Creation?} & \textbf{Attended study 1}& \textbf{Attended study 2} \\ \hline
E1   & More than 5 years & Once every few months & Commercial video & Yes & Yes & Yes
\\ \hline
E2       &  More than 5 years & Once every few months & Animation video, Film & Yes& Yes & Yes
\\ \hline
E3      & 2-5 years  & Once every few months & Film & No & Yes & Yes \\ \hline
E4      &  More than 5 years & Monthly & Vlog, Knowledge sharing video& Yes & Yes & No  \\ \hline
\bottomrule
\end{tabularx}
\end{table}

\begin{table}[ht]
\centering
\small
\caption{User Study Survey Results: Participants were asked a series of 5-point Likert-scale questions about their experience. They rated their sense of creative freedom and transparency when using the system as well as how the AI automations impacted their productivity and creativity. They also rated utility of the structures in helping them complete various tasks.} 
\begin{tabular}{llcccc}
\toprule
\hline
{\multirow{2}{*}{\textbf{Category}}}   &
{\multirow{2}{*}{\textbf{Factor}}} & 
\multicolumn{2}{c}{Novices}     & \multicolumn{2}{c}{Experts}\\
\cmidrule(lr){3-4}\cmidrule(lr){5-6}
  & & Mean & SD & Mean & SD \\
\hline
{\multirow{2}{*}{\textbf{Overall}}}
& Creative Freedom  & 4.67 & 0.52 & 4 & 0.82\\
& Transparency & 4.33 & 0.52 & 3.5 & 1\\ 
\hline
\multicolumn{6}{c}{\textbf{Automations}} \\
\hline
\textbf{Synchronize} & Productivity  & 5 & 0 & 5 & 0\\
\textbf{Functions} & Creativity  & 3.33 & 0.82 & 3.5 & 0.58\\
\hline
\textbf{Generative} & Productivity  & 4.83 & 0.41 & 5 & 0\\
\textbf{Functions} & Creativity  & 4.5 & 0.84 & 3.5 & 0.96\\
\toprule
\multicolumn{6}{c}{\textbf{Structure Utility}} \\
\hline
\multicolumn{1}{l}{\multirow{2}{*}{\textbf{Canvas}}}
& Brainstorming  & 4 & 0.89 & 3.5 & 1.29\\
& Asset Organization & 3.83  & 1.33 & 4 & 0.82\\ 
\hline
\multicolumn{1}{l}{\multirow{2}{*}{\textbf{Narrative Editor}}}
& Brainstorming  & 3 & 1.27 & 4.25 & 0.96\\
& Narrative Development & 3.83  & 1.83 & 4.5 & 0.58\\
\hline
\multicolumn{1}{l}{\multirow{3}{*}{\textbf{Scene Planner}}}
& Brainstorming  & 4.16 & 0.98 & 2.75 & 0.96\\
& Narrative Developmen & 4.5 & 1.22 & 4.25 & 0.5\\
& Scene Planning & 5 & 0 & 5 & 0\\
\hline
\multicolumn{1}{l}{\multirow{4}{*}{\textbf{Timeline Editor}}}
& Narrative Developmen & 3.17 & 1.72 & 2.25 & 1.26\\
& Scene Planning & 3.5 & 1.76 & 2.25 & 1.26\\ 
& Previewing Results & 5 & 0 & 5 & 0\\
& Pacing Adjustment & 4.17 & 1.33 & 5 & 0\\
\hline
\bottomrule
\end{tabular}

\label{tab:study_result}
\end{table}


\newpage
\subsection{Literature Analysis Results by Domain }
\label{sec:apx-lit}
\begin{table}[htbp]
\centering
\caption{Compositional Structures in Different Content Creation Domains}
\label{table:comp_table}
\small
\begin{tabularx}{\textwidth}{>{\centering\arraybackslash}m{3cm} | >{\centering\arraybackslash}m{1.85cm} | >{\centering\arraybackslash}m{1.5cm} | >{\centering\arraybackslash}m{2.7cm} |>{\centering\arraybackslash}m{1.8cm} | >{\raggedright\arraybackslash}m{5cm}}
\toprule
\textbf{Domain} & \textbf{Compositional Structures} & \textbf{Content Aspects} & \textbf{Function} & \textbf{Prior Works} & \textbf{Design Decisions Revolving Around Compositional Structures} \\ 
\toprule
\hline



\multirow{4}{=}{\centering\textbf{Writing} (Creative, Argumentative, Academic)}& Canvas & Narrative & Material organization, Ideation, Notetaking & \cite{Passage,inkplanner}  & 
\multirow{4}{=} {\parbox{5cm} {
\vspace{6px}
``organize argument structures through synchronized text editing and visual programming” \cite{visar} 
\vspace{6px}
\newline 
``integrate multiple prewriting strategies into an iterative and flexible workflow” \cite{inkplanner} 
\vspace{6px}
\newline ``we focused our design on facilitating the transfer of information across applications while tracking its provenance” \cite{Passage}}}
\\
\cline{2-5}
& 
Narrative Graph & Narrative & Narrative exploration and development & \cite{kim2023metaphorian, visar, inkplanner, talebrush} &
\\
\cline{2-5}
& Text Editor   & Narrative, Spatial & Narrative development, Production  & \cite{microsoft_word_2024, Passage} & \\
\cline{2-5}
& Layout Editor & Spatial            &  Layout exploration and generation  & \cite{beyondgrids, AdaptiveLayoutDocuments, adobe_indesign} & \\
\hline

\multirow{2}{=}{\centering\textbf{Podcast}} &  Time-aligned Transcript & 
Narrative, Temporal &  Transcript-based clip editing (Rough Cut) 
& \cite{underscore, contentbased, semanticSpeechEditing} & 
\multirow{4}{=} {\parbox{5cm} { 
``edit the transcript directly using standard word processing `cut and paste' operations, which extract the corresponding time-aligned speech” \cite{semanticSpeechEditing} 
}}
\\
\cline{2-5}
&
\parbox{1.8cm}{\centering \strut\newline Timeline\newline \strut } & 
\parbox{1.6cm}{\centering \strut\newline  Temporal \newline \strut } & 
\parbox{2.8cm}{\centering \strut\newline Timeline-based clip editing (Fine Cut)  \newline \strut  } 
& \cite{descript2024, adobe_audition}  & 
 \\
\hline


\multirow{4}{=}{\centering\textbf{Music}}
&  Canvas &  Narrative
&  Material organization, Ideation, Notetaking  & \cite{challengeinmusicscorewriting, paperSubstrateForMusic, musicartistcaptureideas} & 
\multirow{4}{=} {\parbox{5cm} {
``Composers create their own individual ad hoc strategies for expressing ideas, and often move back and forth between multiple representations " ~\cite{polyphony}
\vspace{6px}
\newline
``(Canvas) offering composers the freedom to arrange scores and musical fragments spatially, adapting the layout to the specific task at hand. " ~\cite{challengeinmusicscorewriting}


}} \\
\cline{2-5}
 &  Tone-Network &  Congruent  &   Chord composition  &  \cite{PaperTonnetz} & \\
\cline{2-5}
& Staff  &  Congruent, Temporal &   Notation-based composition    & \cite {paperSubstrateForMusic, musink} & \\
\cline{2-5}
&  Timeline & Congruent, Temporal & Production  & \cite{apple_logic_pro, ableton_website, audacity_website} & \\
\hline


\multirow{5}{=}{\centering\textbf{Interactive Text-Visual Narrative} (Interactive Article and Comics)}
& Canvas & Spatial, Narrative &  Material organization, Ideation, Sketching  & \cite{DataToon, StickyLand}  & 
\multirow{5}{=} {\parbox{5cm} {
``making the correspondence explicit and consistent is essential when presenting multiple representations…  the mapping between code, story, and comic should be clear” ~\cite{suh2020coding}
\vspace{6px}
\newline
``a section of a narrative and its corresponding visuals will be organized within one block throughout the entire design process to enable flexible prototyping ... while ensuring their correspondence" ~\cite{cao2023dataparticles}

}} \\
\cline{2-5}
 & Panel-based Editor &  Narrative, Spatial & Narrative and layout development & \cite{datacomicspattern, suh2020coding, codetoon, ToonNote} & \\
\cline{2-5}
 & Section-based Editor &  Narrative, Spatial & Narrative and layout development & \cite{cao2023dataparticles, vizflow} & \\
\cline{2-5}
 & Notebook-based Editor &  Narrative, Spatial & Narrative development, Visual creation & \cite{ToonNote, slide4n, outlinespark} &\\
 \cline{2-5}
 & Enhanced Text Editor & Narrative, Spatial & Narrative development, Interaction creation & \cite{crossdata, idyllstudio, livingPapers} &\\
\hline



\multirow{4}{=}{\centering\textbf{Video}}
& \parbox{1.8cm}{\centering \strut\newline  Media Gallery  \newline \strut } &\parbox{2.3cm}{\centering \strut\newline - \newline \strut }  & \parbox{2.8cm}{\centering \strut\newline Media organization  \newline \strut } & \cite{lave,adobe_premiere} & 
\multirow{4}{=} {\parbox{5cm} {
\vspace{2px}
``Establishing elastic and customized mappings between animation and performance to enable graphic elements to adapt to real-time speech and gestures to achieve synchronization and expressive presentation effects.” ~\cite{cao2024elastica}
\vspace{6px}
\newline
“enabling synchronized browsing of the captions, script and summary for easy access to details or context at any point in the film” ~\cite{SceneSkim}}
} \\
\cline{2-5}
& Storyboard  & Narrative, Spatial &  Narrative development, Asset arrangement.  & \cite{videoMosaic, henricksonStoryboard, hart2013art, storeoboard}& \\
\cline{2-5}
&  Time-aligned Transcript & Narrative, Temporal  &  Transcript-based clip editing (Rought Cut) & \cite{truong2016quickcut, crosscast, descript2024, dataplayer} & \\
\cline{2-5}
&  Timeline & Congruent, Temporal  &  Timeline-based clip editing (Fine Cut)  & \cite{adobe_premiere, descript2024, lave} & \\

\hline
 \bottomrule
\end{tabularx}
\end{table}


\newpage
\subsection{Implementation of AI Integration in \system: Generation Details and Prompts  }
\label{sec:apx-prompt}



\begin{table}[htbp]
\centering
\caption{AI Functionalities within \textbf{Canvas}}
\label{table:prompt-canvas}
\small
\begin{tabularx}{\textwidth}{>{\centering\arraybackslash}m{2.8cm}|>{\centering\arraybackslash}m{1.8cm}|>{\centering\arraybackslash}m{3.5cm}|>{\raggedright\arraybackslash}m{8.6cm}}
\toprule
\hline
\textbf{AI Feature} & \textbf{Context} & \textbf{Generation Details} & \textbf{Prompt Used for Generation} \\ \hline
Generate notes based on the prompt  & Parent Asset node & Parse and query document using OpenAI Assistant API      & \textit{\pt{prompt}. make your response in the most concise way possible. }
\\ \hline
Generate description/caption for images    &  Image in the Asset node    &  Query image using OpenAI API GPT4-visual-preview &  \textit{Describe the visual scene in the image to a filmmaker in a concise way. Consider shot type and cinematic style. Make your response as short and concise as possible. Only use 1 sentence. }
\\ \hline
Regenerate image based
on revised prompt      &  Image in the Asset node    & Generate image with prompt using OpenAI API Dall-E-3      & \textit{Generate an image based on the prompt exactly. do not change or revised prompt for generation: \pt{prompt} }     \\ \hline
\bottomrule
\end{tabularx}
\end{table}


\begin{table}[hbp]
\centering
\caption{AI Functionalities within \textbf{Narrative Editor}}
\label{table:prompt-editor}
\small
\begin{tabularx}{\textwidth}{>{\centering\arraybackslash}m{2.6cm}|>{\centering\arraybackslash}m{2cm}|>{\centering\arraybackslash}m{3.4cm}|>{\raggedright\arraybackslash}m{8.6cm}}
\toprule
\hline
\textbf{AI Feature} & \textbf{Context} & \textbf{Generation Details} & \textbf{Prompt Used for Generation} \\ \hline
Generate talking points within a section     & Section heading     & Generate text using OpenAI GPT-4 chat completions API $\rightarrow$ parse the coded response   & 
\textit{you are a video creator for a video about for-
tune cookie origin. You need to come up with
compelling narratives and visuals for the video
planning. You task is generate several talking
points within the given section. The talking
points should present a narrative that fits into
the section. The talking points should commu-
nicate what you want to deliver in each scene.
Each talking point should be one sentence long.
Talking points should have specific details. sec-
tion: \pt{content} response with a string that
contains 2-4 concise and informative talking
points. The talking points should flow logically.
They should not repeat each other. separate
each talking point with \#\#\#. don’t index them,
don't add quotes or prefix.}
\\ \hline
Generate talking points / sections   &  Assets, notes in canvas and existing content in the editor  & Generate text in context
of document using OpenAI
Assistant threads API      & \textit{Consider the video creation prompt: \pt{query}.
The video should be according to the content in
the file. Give me the possible section headings
and talking points that would be in such a video
and that would form a cohesive narrative. Each
heading should be a short sentence. Each talk-
ing point should be one sentence long. Talking
points should have specific details. The talking
points should flow logically. They should not
repeat each other. Give me 2 section headings
and 2-3 concise and informative talking points
within each section. Give me the response as
one string. Each section heading should be pre-
fixed by \%\% and each talking point should be
prefixed by \#\#. First, give me the section heading
and talking points of the first section. Then give
me the section heading and talking points of the
second section. Make your response in the most
concise way possible. Don’t include sources. Ex-
ample response: \%\%Section 1\#\#Talking point
1\#\#Talking point 2\%\%Section 2\#\#Talking point
3\#\#Talking point 4}   \\ \hline
\bottomrule
\end{tabularx}
\end{table}


\begin{table}[htbp]
\centering
\caption{AI Functionalities within \textbf{Scene Planner Grid}}
\label{table:prompt-grid}
\small
\begin{tabularx}{\textwidth}{>{\centering\arraybackslash}m{2.4cm}|>{\centering\arraybackslash}m{2.4cm}|>{\centering\arraybackslash}m{4cm}|>{\raggedright\arraybackslash}m{7.3cm}}
\toprule
\hline
\textbf{AI Feature} & \textbf{Context} & \textbf{Generation Details} & \textbf{Propmt Used for Generation} \\ \hline
Refine script   &  Cells in the same row and column   & Generate text using OpenAI GPT-4 chat completions API   & 
\textit{You are video producer. Your task is to refine text
be a voice-over. Make it as engaging, direct and concise as possible. text to refine: \pt{content}  }  \\ \hline

Suggest split of script based on Scene  &   \nul  & Generate text using OpenAI GPT-4 chat completions API
$\rightarrow$ Parse the coded response  & \textit{You are a professional video producer, trying to segment the script. Separate the text narrative delimited by triple backticks into segments.
Different segments can be the result of different accompanying visuals or changing in subject.
Consider breaking sentences into multiple segments if the sentence can be represented by different visuals.
Format the response as a list of strings (where the strings are substrings of the given text narrative).
Make your response as short and concise as possible.
text description: \pt{script}}
\\ \hline

Visual description generation  &  Visual style, script, and visual descriptions of scenes in the same
storyline paragraph  & Generate text using OpenAI GPT-4 chat completions API
& 
\textit{
You are a professional video producer, trying to couple the visual with the script. Describe the visual you will use based on the list of text narrative segments and accompanying visuals delimited by triple backticks and intended visual style.
The following is a list of dictionaries with 2 keys: visual description and script segment.
List of segments and visual descriptions: \pt{script}
Overall visual style: \pt{style} (ignore if N/A)
You want to describe a static visual scene for the list item where the visual description field says "PROVIDE".
Consider the shot type and visual style.
Format your response as a string that describes a visual scene that can be pictured with one image.
Make your response as short and concise as possible. Only use 1 sentence.}
\\ \hline

Visual preview generation  &   Script, generation style, all visual descriptions, storyline paragraph & If no visual description, generates visual description $\rightarrow$ generate image using OpenAI image generation with Dall-E-3 $\rightarrow$ Parse the coded response  & \textit{\pt{description}, in \pt{style} aesthetic.}
\\ \hline
\bottomrule
\end{tabularx}
\end{table}


\begin{table}[ht]
\centering
\caption{AI Functionalities within \textbf{Timeline Editor}}
\label{table:prompt-timeline}
\small
\begin{tabularx}{\textwidth}{>{\centering\arraybackslash}m{2.5cm}|>{\centering\arraybackslash}m{2cm}|>{\centering\arraybackslash}m{4.6cm}|>{\raggedright\arraybackslash}m{6.6cm}}
\toprule
\hline

\textbf{AI Feature} & \textbf{Context} & \textbf{Generation Details} & \textbf{Propmt Used for Generation} \\ \hline

Align imported audio file with the existing time segments in the audio track &  \nul &  Generate transcription of the audio with timestamps using OpenAI Whisper-1 $\rightarrow$ segment the transcription text to align with the script column in the grid $\rightarrow$ parse coded response & 

\textit{Segment the audio transcription according to the original script segments. Original script segments: \pt{original segments} transcription: \pt{transcript} Return the transcription with \#\#\# in the places where you plan to split the transcription into segments.
}

\\ 
\hline
\bottomrule
\end{tabularx}
\end{table}


\begin{table}[ht]
\centering
\caption{AI Functionalities \textbf{across Structures}} 
\label{table:prompt-across}
\small
\begin{tabularx}{\textwidth}{>{\centering\arraybackslash}m{1.6cm}|>{\centering\arraybackslash}m{2.6cm}|>{\centering\arraybackslash}m{2.5cm}|>{\centering\arraybackslash}m{2.7cm}|>{\raggedright\arraybackslash}m{6.3cm}}
\toprule
\hline

\textbf{Structures} & \textbf{AI Feature} & \textbf{Context} & \textbf{Generation Details} & \textbf{Propmt Used for Generation} \\ \hline
N $\Rightarrow$ C & Generate notes based on talking points & Parent Asset and all narrative content within a section & Parse and query document using OpenAI Assistant API      & \textit{Getting relevant content for \{talking point\}} \\ \hline

N|S $\Rightarrow$ C & Finding relevant notes while writing script & narrative content being edited &\multicolumn{2}{>{\raggedright\arraybackslash}m{8.8cm}}{Create embeddings for notes and talking point using OpenAI API text-embeddings-3-small $\rightarrow$ Get notes whose embeddings are less than 0.93 distance apart from the talking point embedding} \\ \hline

C $\Rightarrow$ N & Form talking points with note & All content in the editor & Generate text with OpenAI GPT-4 chat completions API & 
\textit{Your job is to revise the current content with the note, make it fit into the existing narrative. the current content is part of the exiting narrative. You need to understand where current content is, and how to make it more solid with the note and how the revised version can smoothly fit into the narrative flow. current content: \pt{current content}; note: \pt{note content}  
existing narrative: \pt{talking points}. Your generated content should be as direct and concise as possible. one sentence.}
\\ \hline

N $\Rightarrow$ S & Populate storyline column to the relevant script/visual columns & \nul & Generate text with OpenAI GPT-4 chat completions API with few-shot prompting * & 
\textit{Your task is to segment the storyline content into voicee
over and visual description. You will be provided with the sandbox content. response a dictionary of the segmented result." sandbox content: \pt{content} only use the content provided, don’t add new content!}
\\ \hline

T $\Rightarrow$ S & Fine-tune script by adjusting the time segments & Existing script in Scene Planner & Generate text using OpenAI GPT-4 chat completions with
few-shot prompting $\rightarrow$ parse coded response & 
\textit{You are a wordsmith. Make the text \pt{length}
words \pt{shorter | longer}. Do not change the
meaning of the sentence, but you can add or
remove words. The output sentence should
be meaningful and cohesive. Text: \pt{content}
Give the original text with annotations. Put \#\#\#
around the words that you added to the original
sentence. Put 
removed from the original sentence. Respond
with only the original text with annotations. Do
NOT prefix the response with anything.}
\\ \hline

S $\Rightarrow$ T  & Generate audio voice
over based on script & \nul &
\multicolumn{2}{>{\raggedright\arraybackslash}m{8.8cm}}{Generate audio voice
over based on script \&
Generate audio file using
OpenAI Text-To-Speech
API -> generate transcrip-
tion of the new audio
file to get timestamps
for words using OpenAI
Whisper-1}\\
\hline
\bottomrule

\end{tabularx}
\end{table}